\documentclass[11pt,a4paper,oneside]{article}
\usepackage[top=3cm, bottom=3cm, left=2cm, right=2cm]{geometry}
\usepackage[english]{babel}
\usepackage[utf8]{inputenc}
\usepackage[T1]{fontenc}
\usepackage{lmodern}          % modern font with accents 

\usepackage{amsthm,amsmath,amssymb,mathrsfs,dsfont,mathtools}

\usepackage{bm}
\usepackage{graphicx}
\usepackage[colorinlistoftodos]{todonotes}
\usepackage[colorlinks=true, allcolors=blue]{hyperref}
\usepackage{booktabs,subcaption}
\usepackage{hyperref}
\usepackage[all]{nowidow}
\usepackage{setspace}
\usepackage{algorithm}
\usepackage[noend]{algpseudocode}
\usepackage{tikz} % For the plot
\usepackage[sf]{titlesec}
\usepackage{sectsty}
\allsectionsfont{\centering \normalfont\scshape} % Make all sections centered, the default font and small caps
\usepackage{lipsum}
\usepackage{adjustbox}
\usepackage{lineno} 
\usepackage{enumitem} 
\usepackage{natbib}

\usepackage{amssymb}   % for \checkmark
\usepackage{pifont}    % for \ding

\newcommand{\cmark}{\checkmark}
\newcommand{\xmark}{\ding{55}}

\usepackage{authblk}
\title{
Bayesian modeling of predictive stability under generalized Arrhenius non-linear kinetics
}
\date{\today}
\author[1]{Lorenzo Mauri}
\author[2]{Samantha Gleich}
\author[2]{Michael Johnson}
\author[2]{Andy Liaw}
\author[2]{Daniel Skomski}
\author[2]{Federico Ferrari}

\affil[1]{Department of Statistical Science, Duke University, Durham, NC, USA}
\affil[2]{Merck \& Co., Inc., Rahway, NJ, USA}

\theoremstyle{definition}

\usepackage[resetlabels]{multibib}
\newcites{Supp}{References}

\usepackage{amsmath}
\usepackage{newtxtext}
\usepackage[subscriptcorrection]{newtxmath}

\usepackage{natbib}

\graphicspath{{./fig/}}

\usepackage{color}

\usepackage{hyperref}

\usepackage{graphicx}
\graphicspath{{fig/}}
\usepackage{caption}
\usepackage{subcaption}
\usepackage{dirtytalk}
\usepackage{hyperref}

\usepackage{subcaption}

\usepackage{comment}

\usepackage{algorithm}
\usepackage{algpseudocode}
\usepackage[T1]{fontenc}  % in the preamble, to support accented characters properly

\begin{document}

\maketitle

\begin{abstract}

Predictive stability modeling is widely used in pharmaceutical development to establish product shelf-life by identifying the kinetic parameters that govern degradation behavior and using these parameters to extrapolate out in time. Recent approaches typically assume linear degradation dynamics and rely on the Arrhenius equation to model temperature-specific reaction rates. However, observed degradation profiles may reflect nonlinear reaction dynamics and more complex temperature-specific reaction rates, potentially leading to reduced accuracy in long-term predictions and motivating a more flexible approach. 
We propose a novel Bayesian framework which extends predictive stability modeling beyond simple Arrhenius low-order kinetic dynamics and accommodates complex degradation mechanisms.
The degradation process is modeled through the Sestak–Berggren kinetic formulation, a flexible ordinary differential equation encompassing many classical degradation models as special cases. To facilitate model interpretability and model exploration, probabilistic model selection on kinetic parameters automatically determines whether nonlinear reaction components are supported by the data, allowing the model to adaptively collapse to simpler dynamics when appropriate.
To address departure from classical Arrhenius kinetics, we develop a generalized temperature reaction rate formulation in which temperature-specific reaction rates are shrunk toward Arrhenius predictions through a hierarchical shrinkage prior while allowing departures when supported by the data. This approach yields robustness to misspecification and reduces sensitivity to potentially influential high-temperature observations commonly encountered in accelerated stability experiments.
We demonstrate the usefulness of the proposed approach in two stability studies. The resulting framework provides a principled and flexible statistical tool for model-informed shelf-life determination.

\end{abstract}

%\begin{keywords} 
%Predictive stability;
%\end{keywords}

\section{Introduction}

Real-time stability studies are conducted in the pharmaceutical industry to verify that a drug substance or drug product remains effective and safe through the set shelf-life. These real-time stability studies typically involve incubating the drug at the anticipated storage temperature and testing the drug periodically for purity and potency to confirm that critical quality attributes (CQAs) remain within set specification limits through shelf-life \citep{bajaj2012, chau2023}. 
Standard approaches for running these real-time stability studies are outlined in the International Council for Harmonisation (ICH) stability guidelines and denote that data used to support the shelf-life claims for biotherapeutic compounds should cover the duration of the proposed shelf-life (ICH Q1A(R2) 2003). Due to their inherent long-term nature, such real‑time stability studies rank among the most time‑ and resource‑intensive activities in pharmaceutical development and often lead to delays in setting clinical or commercial shelf-life \citep{kuzman2021, chau2023}.

The latest revision of the ICH stability guidelines (Q1, Annex 2) addresses the bottleneck posed by real‑time stability studies by endorsing enhanced stability‑modeling approaches for establishing the shelf-life of drug substances and drug products (ICH Q1 Draft, 2025). These enhanced stability modeling approaches typically use data collected at storage, accelerated, and stressed conditions (i.e., accelerated stability data) to train an Arrhenius‑based kinetic model that will describe the degradation of a compound over time and across temperatures \citep{bajaj2012}. Several studies have demonstrated that Arrhenius-based kinetic models trained on short-term stability data (e.g., 6 months of data) can accurately predict drug substance or drug product stability through shelf-life (e.g., 36 months;  \citet{clenet2014advanced, kuzman2021, huelsmeyer2023universal, dillon_et_al_24, ferrari2025, skomski_25}). For example, \citep{clenet2014advanced} used 6 months of protein concentration data to train a kinetic model that could predict the protein concentration of a protein-based vaccine with a high degree of accuracy at 2 years. Similarly, \citet{huelsmeyer2023universal} used short-term data and kinetic models to generate accurate long-term predictions for monoclonal antibodies, fusion proteins, vaccines, and other biological drug products.

Although predictive stability modeling has provided direct value in the development of biopharmaceutical products in recent years, it is important to recognize that there are instances where simple, Arrhenius-based kinetic models will fail to fully capture stability profiles over time. For example, one-step, zero- or first-order reaction mechanisms are often assumed when modeling pharmaceutical degradation reactions \citep{bhangare2022}; however, studies have shown that some biotherapeutic degradation reactions may follow higher-order or multi-step degradation pathways \citep{clenet2014advanced}. There have also been instances where Arrhenius-based kinetics do not fully capture temperature-dependent degradation rates \citep{kayser2011, wangroberts2013, bunc2022}, suggesting that alternatives to an Arrhenius-based kinetic model may be required. These outstanding challenges underscore the need for a flexible model-fitting framework that can capture complex degradation mechanisms, while also prioritizing model interpretability.

To address the complex kinetic mechanisms that may contribute to drug stability, while also prioritizing model simplicity, we developed a Bayesian modeling framework that can be used to (1) capture non-linear reaction mechanisms and (2) model deviations from the Arrhenius equation. This framework leverages spike-and-slab priors on the kinetic parameters, to identify nonlinearity in stability data and model that nonlinearity when appropriate, or collapse toward a more parsimonious, less flexible solution. A spike-and-slab prior is a mixture prior combining a component concentrated near zero (the \say{spike}) with a more diffuse component (the \say{slab}). This allows the data to determine whether a model component should be effectively excluded or retained, providing Bayesian model selection within a single fitted model. Our modeling framework also uses a hierarchical shrinkage prior to allow deviations from Arrhenius kinetics, when the data used in model fitting are not well-described by the classical Arrhenius equation. The spike-and-slab and adaptive shrinkage approaches that are integrated into our Bayesian kinetic modeling framework enable data-driven identification of complex degradation pathways from stability data.

\section{Model Formulation}

We propose a Bayesian framework for predictive stability modeling with three key features: (i) a flexible kinetic formulation capturing nonlinear degradation dynamics, (ii) a generalized Arrhenius formulation allowing data-driven deviations from classical temperature–rate relationships, and (iii) a variable selection mechanism that controls model complexity and adaptively recovers simpler models when supported by the data.

\subsection{Sestak-Berggren Model}

We model the observed CQA trajectory as 
\begin{equation}\label{eq:cqa}
    \begin{aligned}
        y_i = Y_0 + (Y_\infty - Y_0) \alpha(t_i; T_i) + \varepsilon_i, \quad \varepsilon_i \sim N(0, \sigma^2),
    \end{aligned}
\end{equation}
where $t_i$ and $T_i$ denote time and temperature (measured in Kelvin) of the $i$-th observation, respectively. $Y_0$ represents the unknown initial state, while $Y_\infty$ is the final value of the CQA when the reaction is completed, which is often known. Here, $\alpha(t, T)$ represents the fraction reacted with $\alpha(\infty; T) = 1$, and $  \frac{d \alpha(t, T)}{d t} \geq 0$.  The latent reaction progress $\alpha(t, T)$ evolves according to the ordinary differential equation (ODE)
\begin{equation} \label{eq:d_alpha}
    \begin{aligned}
       \frac{d \alpha(t, T)}{d t} = k(T) f(\alpha(t; T)).
    \end{aligned}
\end{equation}
In \eqref{eq:d_alpha}, $k(T)$ is the temperature dependent rate constant, which is generally modeled using the Arrhenius equation \citep{atkins2018physical}, but in Section \ref{subsec:generalized_Arrhenius}, we discuss its generalization to handle cases where it might be violated. $f(\alpha(t; T))$ is a kinetic model that describes the underlying reaction mechanism given here by the Sestak-Berggren (SB) Model \citep{sb_model}        
\begin{equation}\label{eq:f_alpha}
    f(\alpha(t; T)) = (1 - \alpha(t; T))^n \alpha(t; T)^m
\end{equation}
 where $m$ is the autocatalytic parameter and $n$ is the reaction order. The SB model is appealing because it encompasses several classical kinetic specifications as special cases while still permitting more general nonlinear behavior when required by the data. 
 Closed-form solutions to \eqref{eq:d_alpha} exists for specific parameter configurations (see Table \ref{tab:special_cases_sb}), but, in general, $\alpha(t; T)$ is defined implicitly via the ODE and computed numerically.

\begin{table}[h!]
\centering
\renewcommand{\arraystretch}{1.5}
\begin{tabular}{@{}cccll@{}}
\toprule
\textbf{$n$} & \textbf{$m$} & \textbf{CQA equation for $Y(t)$} & \textbf{Type} & \textbf{Notes} \\
\midrule
0 & 0 & \( Y_0 + (Y_\infty - Y_0)\, k(T)\, t\) & Zero-th-order$^*$ & Linear growth \\
1 & 0 & \( Y_0 + (Y_\infty - Y_0)\big(1 - e^{-k(T)t}\big) \) & First-order & Exponential growth \\
2 & 0 & \( Y_0 + (Y_\infty - Y_0)\left(1 - \frac{1}{1 + k(T)t}\right) \) & Second-order & Reciprocal linear growth \\
$n$ & 0 & $Y_0 + (Y_\infty - Y_0)
\left[1 - \left\{1 - (1-n)k(T)t\right\}^{\frac{1}{1-n}}\right]$ & $n$-th order & General $m=0$ and $n \in \mathbb N \setminus \{1\}$ form \\
\midrule
1 & 1 & \( Y_0 + (Y_\infty - Y_0)\frac{1}{1 + C(T)e^{-k(T)t}} \) & Logistic & Saturating S-shaped growth \\
\bottomrule
\end{tabular}
\caption{Special cases of the SB model for $Y(t)$ for different values of $(n,m)$. The first block corresponds to $m=0$, while $(n,m)=(1,1)$ yields logistic dynamics, with 
$C(T) = \frac{1 - \alpha_0(T)}{\alpha_0(T)}$, where $\alpha_0(T)=\alpha(0,T)$.  $^*$For the zero-th-order model, $\alpha(t; T) = K(T)t \in [0,1]$ only for $t \in [0, 1/K(T)]$.}
\label{tab:special_cases_sb}
\end{table}

\subsection{Modeling of Kinetic Parameters}
 Most common approaches consider the simplest case for \eqref{eq:f_alpha} setting $m=n=0$, which results in linear dynamics over time, and is often referred to as King–Kung–Fung (KKF) model \citep{king_84}.
 However, empirical evidence often suggests deviation from linearity calling for models that handle more complex scenarios. Therefore, we take a more general approach and consider $m$ and $n$ unknown parameters. We adopt a Bayesian approach and place priors on the kinetic parameters supported on the positive real line.
 A default choice are truncated normal distributions: i.e., 
 \begin{equation}\label{eq:vanilla_priors}
     m \sim TN(0, \sigma_m^2), \quad n \sim  TN(0, \sigma_n^2),
 \end{equation}
 where $TN(0, \tau^2)$ denotes a Normal distribution with mean $0$ and variance $\tau^2$ restricted on the positive part of the real line. Taking a Bayesian approach is particularly advantageous when the data provide limited information about the values of kinetic parameters and obtaining a point estimate can be very challenging, allowing for the uncertainty in the kinetic parameters estimates to be propagated into out-of-sample prediction. 

 Nevertheless, simpler models are often useful due to their ease of interpretation and faster computing times. To this end, we employ a variable selection procedure that shrinks $m$ and/or $n$ toward zero when the data provide limited support for the corresponding kinetic effects. For instance, in some cases, a linear dynamic over time is sufficient to characterize  the degradation mechanism of the molecules well. In these scenarios, a simpler model might be easier to fit (since it does not require one to solve the ODE numerically) and have a better out-of-sample performance due to the lower number of parameters that need to be estimated.
 However, neglecting the autocatalytic or higher order components of the ODE when deviations from linearity are prevalent decreases the predictive accuracy of the models. Scientists do not always know the underlying kinetics of a degradation mechanism in advance, making non-linear model selection a common challenge in applying predictive stability methods. 
 
Motivated by the considerations above, we propose a variable selection procedure to allow $m$ and $n$ deviate from $0$ only when necessary. We do this via a relaxed spike and slab prior \citep{spike_and_slab}, i.e.
\begin{equation}\label{eq:spike_and_slab}
\begin{aligned}
        w \sim (1 - \pi_w)TN(0, \sigma_{\text{spike}, w}^2) &+  \pi_wTN(0, \sigma_{\text{slab}, w}^2),\\
        \pi_w \sim \ &\text{beta}(0.5, 0.5),
\end{aligned} \quad  w \in \{m,n\}.
\end{equation}
\iffalse
The prior in \eqref{eq:spike_and_slab} can be equivalently represented as 
\begin{equation}
\begin{aligned}
        w &\sim \begin{cases}
          TN(0, \sigma_{\text{spike}, w}^2),  \quad \gamma_w = 0,\\
         TN(0, \sigma_{\text{slab}, w}^2), \quad \text{otherwise},
        \end{cases}\\
        \gamma_w &\sim \text{Bernoulli}(\pi_w),\\
        \pi_w &\sim \text{beta}(a, b),\\
\end{aligned} \quad w \in \{m,n\}.
\end{equation}
\fi

Here, $\sigma_{\mathrm{spike},w}^{2}$ is the variance of the narrow spike component, while $\sigma_{\mathrm{slab},w}^{2}$ is the larger variance of the slab component. The spike concentrates on small positive values, encouraging trajectories close to those of the corresponding reduced model, while the slab accommodates larger kinetic effects. Because both components are continuous, this formulation provides approximate model selection through shrinkage, without assigning positive posterior probability to an exact reduced model. 

To summarize posterior support for the mixture components, let $\gamma_w=1$ denote allocation to the slab and $\gamma_w=0$ allocation to the spike. We report the posterior slab-allocation probability $q_w=\Pr(\gamma_w=1\mid\mathbf y)$, estimated by averaging the conditional allocation probabilities $\Pr(\gamma_w=1\mid w,\pi_w)$ over posterior draws.

The prior can also favor other scientifically relevant kinetic specifications. For instance, scientists might believe the dynamics to follow a first order (or exponential growth) model and wonder whether a more general one is necessary. Then, centering the spike component for $n$ at one while retaining the spike for $m$ at zero encourages trajectories close to first-order dynamics.

\subsection{Generalized Arrhenius equation modeling}\label{subsec:generalized_Arrhenius}
The temperature-dependent reaction rate $K(T)$ is commonly modeled using the Arrhenius equation \citep{atkins2018physical}, 
\begin{equation}\label{eq:Arrhenius_eq}
    \begin{aligned}
       k(T) = k_{\text{ref}} \bar g(T), \quad \bar g(T) = \exp\left\{\frac{E_a}{R}\left(\frac{1}{T_{\text{ref}}} - \frac{1}{T}\right)\right\},
    \end{aligned}
\end{equation}
where $k_{\text{ref}}$ is the pre-exponential factor, $E_a$ is the activation energy and $R$ is the universal gas constant, which is equal to $8.314 \ J mol^{-1} K^{-1}$, and $T_{\text{ref}}$ is a reference temperature in Kelvin, which we set to $298K$. Note that \eqref{eq:Arrhenius_eq} imposes a log-linear relationship between inverse temperature and reaction rates. While this structure is widely used in predictive stability modeling, it can be restrictive in the presence of complex biochemical degradation mechanisms. 
In fact, substantial empirical evidence demonstrates systematic deviations from Arrhenius behavior in biochemical systems, particularly in protein aggregation and degradation pathways \citet{kayser2011, wangroberts2013, bunc2022}. In accelerated stability studies, such deviations often arise because elevated temperatures can alter reaction mechanisms, induce phase transitions, or shift the rate-determining step, leading to non-Arrhenius temperature–rate relationships and potentially biased extrapolation \citet{waterman_adami}.  This raises a fundamental modeling question: whether high-temperature observations, which may violate Arrhenius assumptions, should be included in estimation, risking bias due to model misspecification, or excluded, at the cost of increased uncertainty due to reduced sample size.
This might call into question the reliability of predictions obtained using the Arrhenius equation when it is violated and  underscores the need for a more flexible formulation that preserves the inductive bias of Arrhenius kinetics while allowing data-driven deviations when necessary.

Motivated by these considerations, we propose an alternative approach, which we refer as generalized Arrhenius equation (GAE) modeling, where the Arrhenius equation is only used to define a scientifically informed prior, while allowing data-informed deviations from it in estimating the reaction rates. In particular, we model the log-reaction rates as
\begin{equation}\label{eq:prior_gae}
    \log k(T) \mid k_{\text{ref}}, E_a, \sigma_{K_T}^2 \sim N(\log k_{\text{ref}} + \log\{\bar g(T)\}, \sigma_{k_T}^2), \quad (T \in \mathcal T),
\end{equation}
 where $\bar g(\cdot)$ is the defined in \eqref{eq:Arrhenius_eq} and $\mathcal{T}$ is the set of temperatures at which the compound is stored in the study. 
This formulation centers \textit{a priori} the model around the classical Arrhenius relationship while allowing deviations that are supported by the data, effectively inducing partial pooling of temperature-specific reaction rates toward the Arrhenius curve. The prior variance $\sigma_{k_T}^2$ controls the degree of adherence to Arrhenius kinetics at the temperature $T$: small values enforce near Arrhenius behavior, while larger values permit substantial departures, enabling the model to capture non-Arrhenius regimes. To allow data-adaptive shrinkage, we place a hyperprior 
\begin{equation*}
    \sigma_{k_T}^2 \sim \text{Inv-Gamma}\left(a_K, b_K\right),
\end{equation*}
which enables the model to automatically determine the extent of deviation from Arrhenius behavior based on the observed data. In our data analysis, we show that this approach effectively handles violations of \eqref{eq:Arrhenius_eq}, while collapsing on the Arrhenius dynamics when such assumption holds.

As an alternative to modeling deviations from Arrhenius behavior, \citet{chau2023} consider an Arrhenius model with random effects (AE\&RE), in which additional variability is introduced directly at the level of the reaction rates. Their approach augments the Arrhenius formulation with a stochastic term on the log-rate scale, inducing multiplicative variability in $k(T)$ across conditions. Formally, they model the (log) reaction rate for the $i$-th observation as $\log k_i(T) =  \log k_{\text{ref}}  + \log\{\bar g(T)\} + \eta_i$, where $\eta_i \sim N(0, \tau_{\eta}^2)$. 
 This formulation accounts for unexplained heterogeneity in degradation rates, without explicitly modeling systematic departures from Arrhenius kinetics. However, because the random effects are centered around the Arrhenius structure and act as unstructured noise, this approach cannot correct for systematic misspecification of the temperature–rate relationship, leading to biased and unstable extrapolation when Arrhenius assumptions are violated. This is particularly evident in our data analysis presented in Section \ref{subsec:application_gae}.

\subsection{Prior formulation and posterior computation}
We adopt a Normal prior on $Y_0$ and log-normal priors for $k_{\text{ref}}$ and $E_a$, with inverse-gamma hyperpriors on their respective prior variances parameters, i.e.,
\begin{equation*}
    \begin{aligned}
        Y_0 \mid  \tau_{Y_0}^2  & \sim N(\mu_{Y_0}, \tau_{Y_0}^2), \quad &\tau_{Y_0}^2 &\sim IG(a_{Y_0}, b_{Y_0}),\\
         \log k_{\text{ref}}  \mid \mu_{k_{\text{ref}}}, \tau_{k_{\text{ref}}}^2 & \sim N(\mu_{k_{\text{ref}}}, \tau_{k_{\text{ref}}}^2), \quad &\tau_{k_{\text{ref}}}^2 &\sim IG(a_{k_{\text{ref}}}, b_{k_{\text{ref}}}).
    \end{aligned}
\end{equation*}
For the activation energy parameter, we set a log-normal prior $\log E_a  \mid \tau_{E_a}^2 \sim N(\mu_{ E_a }, \tau_{ E_a }^2)$, with a inverse-gamma prior on the prior variance, $\tau_{ E_a }^2 \sim IG(a_{ E_a }, b_{ E_a })$ for the models with Arrhenius dynamics, while we empirically tune $\tau_{E_a}$ for the GAE framework.
We assign an inverse-gamma prior to $\sigma^2$, i.e., $\sigma^2 \sim IG(a_{\sigma}, b_{\sigma})$. Details about hyperparameters specification are reported in the supplementary material. This prior construction naturally extends to hierarchical settings in which measurements are collected across multiple batches and drug configurations. This is particularly important when sample sizes within each group are small, as it enables information sharing while still allowing for differences in group-specific parameters.

Posterior computation is performed via the STAN \citep{stan} implementation of the No-u-Turn Sampler \citep{nuts}. At each iteration of the MCMC, the likelihood computation requires solving for $\alpha(\cdot, \cdot)$ for each point in the dataset, which is approximated numerically via the Dormand-Prince algorithm, a 4th/5th order Runge-Kutta method \citep{dormand_prince}. For all the experiments, we run the sampler for $10000$ iterations discarding the first $2000$ as burn-in, assess convergence using standard diagnostics and via visual inspections of traceplots, which did not reveal convergence issues. A potential extension to speed up computations would be to develop a HMC-within-Gibbs sampler, exploiting the  tractability of full conditionals of some parameters, especially in the hierarchical formulation.

\section{Numerical experiments}\label{sec:numerical_experiments}
In this section, we explore the performance of competing methods across a range of numerical experiments. We are mainly interested in testing the two main model developments, model selection for kinetic parameters and generalized Arrhenius equation modeling, separately.  

The first study focuses on the degradation mechanism: data are generated under the classical Arrhenius equation, while the kinetic parameters $(m,n)$ are varied to represent linear, first-order nonlinear, and autocatalytic nonlinear dynamics. This experiment evaluates whether the Sestak--Berggren formulation improves prediction under nonlinear degradation and whether the spike-and-slab prior can identify unnecessary kinetic terms. The second study focuses on the temperature--rate relationship: data are generated either under the classical Arrhenius equation or under systematic deviations from it at the lowest and highest temperatures. This experiment evaluates the efficiency of Arrhenius-based models when the physical assumption is correct and the robustness of the generalized Arrhenius formulation when it is violated.

Across simulation replicates, we fit the competing models using the training observations and evaluate prediction at held-out $5^\circ\mathrm{C}$ observations, corresponding to the storage condition of primary scientific interest. Predictive accuracy is measured by root mean squared error (RMSE) using posterior predictive means. Uncertainty quantification is assessed by the empirical coverage and average length of $95\%$ posterior predictive intervals for held-out CQA values.

\subsection{(Non)linear dynamics and variable selection}\label{subsec:numerical_experiments_vs}
First, for all the models we use the Arrhenius equation (AE) to model reaction rates and we compare the standard King--Kung--Fung (KKF) model, which assumes linear degradation dynamics, to the Sestak--Berggren (SB) model, which allows nonlinear kinetics. We also study whether the SB model with spike-and-slab priors \eqref{eq:spike_and_slab}, denoted by SB-SAS, can adaptively remove unnecessary kinetic terms.

Data were generated under the Arrhenius equation using the Sestak--Berggren model in \eqref{eq:cqa}--\eqref{eq:d_alpha}, with temperature-specific rates
\begin{equation*}
k(T)=k_{\text{ref}}\exp\left\{\frac{\exp(11.5)}{8.314}\left(\frac{1}{298.15}-\frac{1}{T}\right)\right\}.
\end{equation*}
We considered time points $t \in \mathcal T_t$ and temperatures $T \in \mathcal T_T$, where
\begin{equation*}
\mathcal T_t=\{0,0.25, 0.5,1,2,3,4,5,6,9,12,18,24,30,36\}, 
\qquad 
\mathcal T_T=\{5,15,25,37,40\}^\circ\mathrm{C},
\end{equation*}
with temperatures converted to Kelvin. To mimic an accelerated stability study with sparse long-term measurements at higher temperatures, we retained the full time grid at $5^\circ\mathrm{C}$, the storage condition often of interest, retained observations up to 6 months at $15^\circ\mathrm{C}$, $25^\circ\mathrm{C}$, and $37^\circ\mathrm{C}$, and retained observations only up to 1 month at $40^\circ\mathrm{C}$. For each retained temperature--time pair, three observations were generated. We fixed $Y_0=10$, $Y_\infty=100$, $ k_{\text{ref}}=0.05$, and generated responses from
\begin{equation*}
y_i \sim N\{Y_0+(Y_\infty-Y_0)\alpha(t_i;T_i),\sigma_y^2\}, \qquad \sigma_y=2,
\end{equation*}
where $\alpha(t;T)$ solves the Sestak--Berggren ODE with initial value $\alpha(0;T)=10^{-4}$. We considered three kinetic regimes, $(m,n)=(0,0)$, $(m,n)=(0,1)$, and $(m,n)=(0.1,1)$, corresponding respectively to linear, first-order nonlinear, and autocatalytic nonlinear dynamics. This setup isolates the variable-selection problem, since the Arrhenius assumption is correctly specified and the only source of model complexity is the kinetic form.

First, we examine estimation of the kinetic parameters and posterior allocation to the slab components (Table~\ref{tab:vs_selection}). Under linear dynamics, SB-SAS yields posterior means of $m$ and $n$ closer to zero than the baseline SB model, together with low posterior slab-allocation probabilities. Under first-order dynamics, $(m,n)=(0,1)$, SB-SAS estimates $n$ close to one while shrinking $m$ toward zero. In the autocatalytic scenario, $(m,n)=(0.1,1)$, the posterior means remain close to the generating values, although posterior support for $m$ is more variable, as expected because the autocatalytic component is weaker and harder to separate from the reaction-order effect. These results show that the prior encourages approximately simpler dynamics when appropriate while accommodating nonzero kinetic effects.

\begin{table}[ht]
\centering
\resizebox{\textwidth}{!}{%
\begin{tabular}{lcc|cccc}
\toprule
& \multicolumn{2}{c|}{SB}
& \multicolumn{4}{c}{SB-SAS} \\
\cmidrule(lr){2-3} \cmidrule(lr){4-7}
Scenario
& $\widehat{m}$ & $\widehat{n}$
& $\widehat{m}$ & $\widehat{n}$
& $\widehat{q}_m$ & $\widehat{q}_n$ \\
\midrule
$m = 0,\ n = 0$
& $0.042$ $(0.020)$ & $0.084$ $(0.037)$
& $0.007$ $(0.019)$ & $0.017$ $(0.028)$
& $0.089$ $(0.236)$ & $0.141$ $(0.266)$ \\

$m = 0,\ n = 1$
& $0.048$ $(0.024)$ & $1.120$ $(0.102)$
& $0.020$ $(0.052)$ & $1.043$ $(0.131)$
& $0.116$ $(0.284)$ & $1.000$ $(0.000)$ \\

$m = 0.1,\ n = 1$
& $0.101$ $(0.044)$ & $1.001$ $(0.165)$
& $0.098$ $(0.082)$ & $0.966$ $(0.232)$
& $0.707$ $(0.442)$ & $1.000$ $(0.000)$ \\
\bottomrule
\end{tabular}%
}
\caption{Across-replicate mean (standard deviation) of posterior means of the
kinetic parameters under the baseline SB prior \eqref{eq:vanilla_priors} and the SB-SAS prior \eqref{eq:spike_and_slab}. For SB-SAS, $\widehat q_w$ denotes the Monte Carlo estimate of the posterior slab-allocation probability $q_w=\Pr(\gamma_w=1\mid\mathbf y)$, for $w\in \{m,n\}$. }
\label{tab:vs_selection}
\end{table}

\begin{figure}[ht]
    \centering
    \begin{subfigure}{0.33\textwidth}
        \centering
        \includegraphics[width=\linewidth]{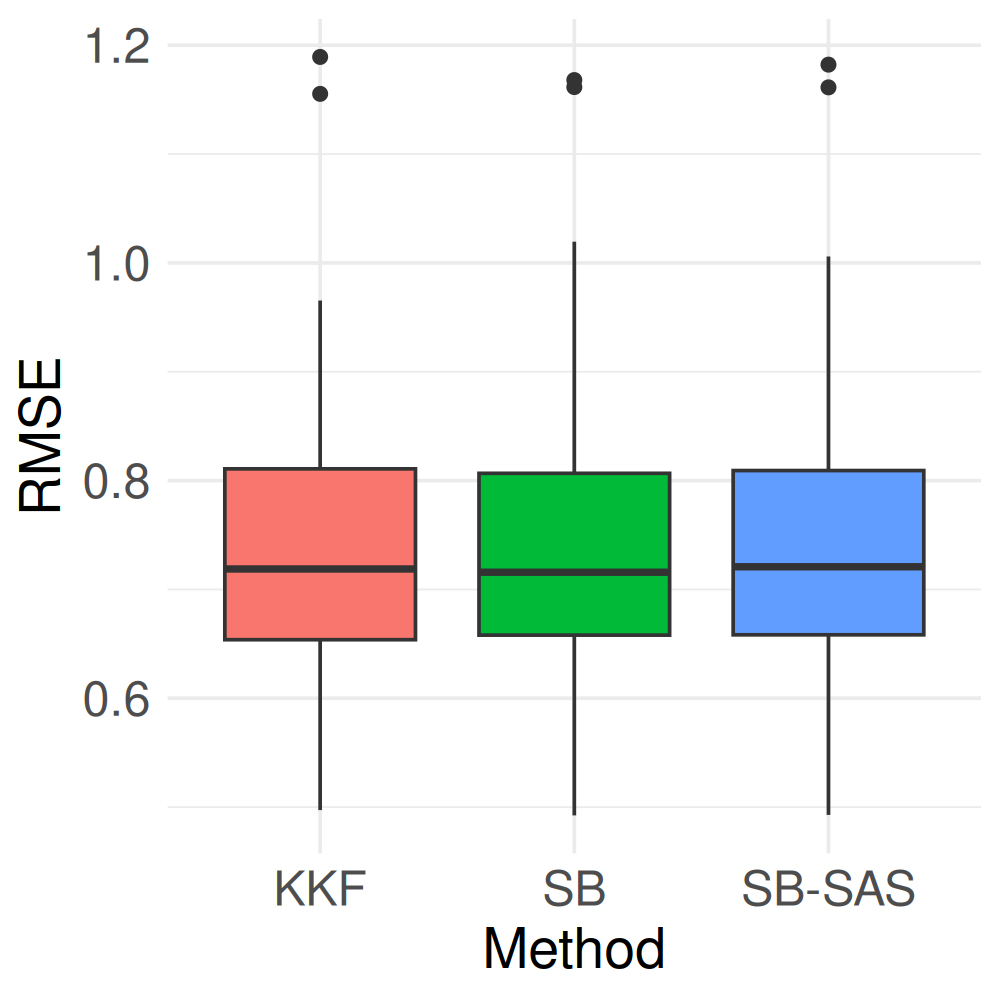}
        \caption*{$m=0$ \& $n=0$}
    \end{subfigure}
    \hfill
    \begin{subfigure}{0.33\textwidth}
        \centering
        \includegraphics[width=\linewidth]{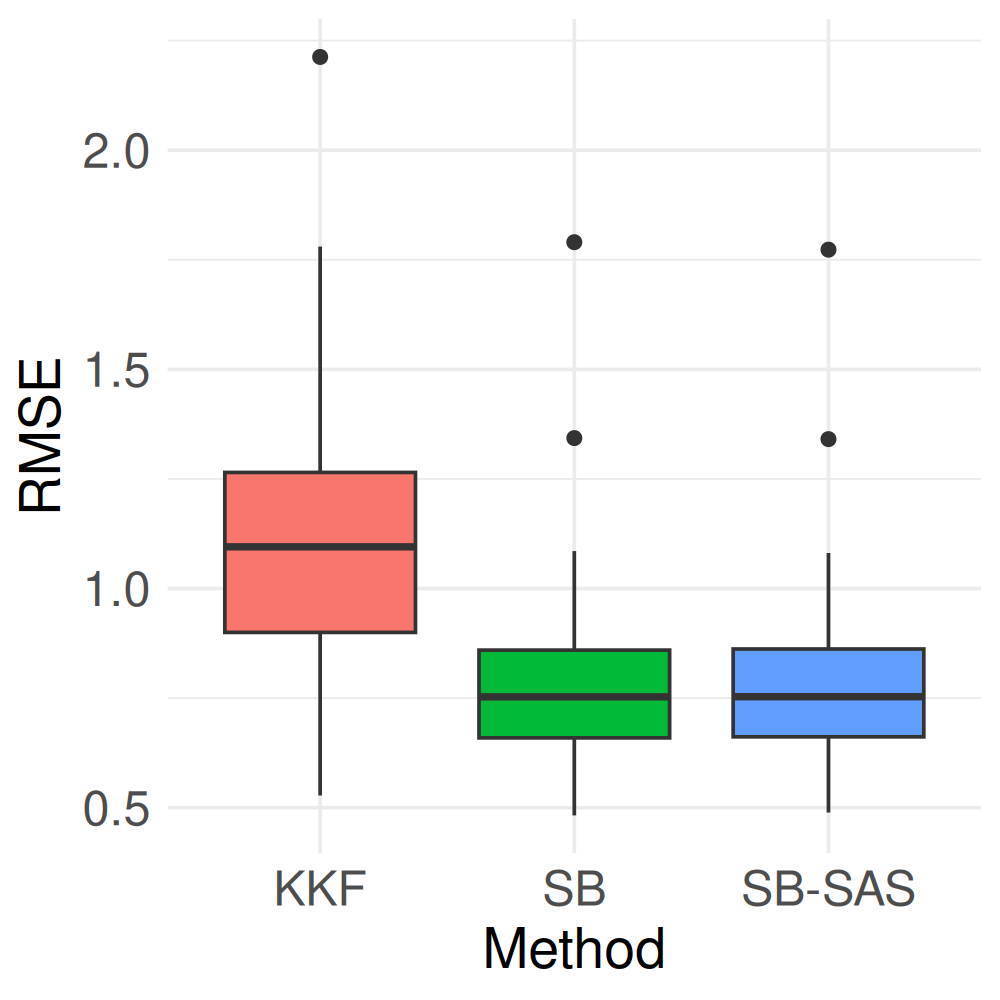}
        \caption*{$m=0$ \& $n=1$}
    \end{subfigure}
    \hfill
    \begin{subfigure}{0.33\textwidth}
        \centering
        \includegraphics[width=\linewidth]{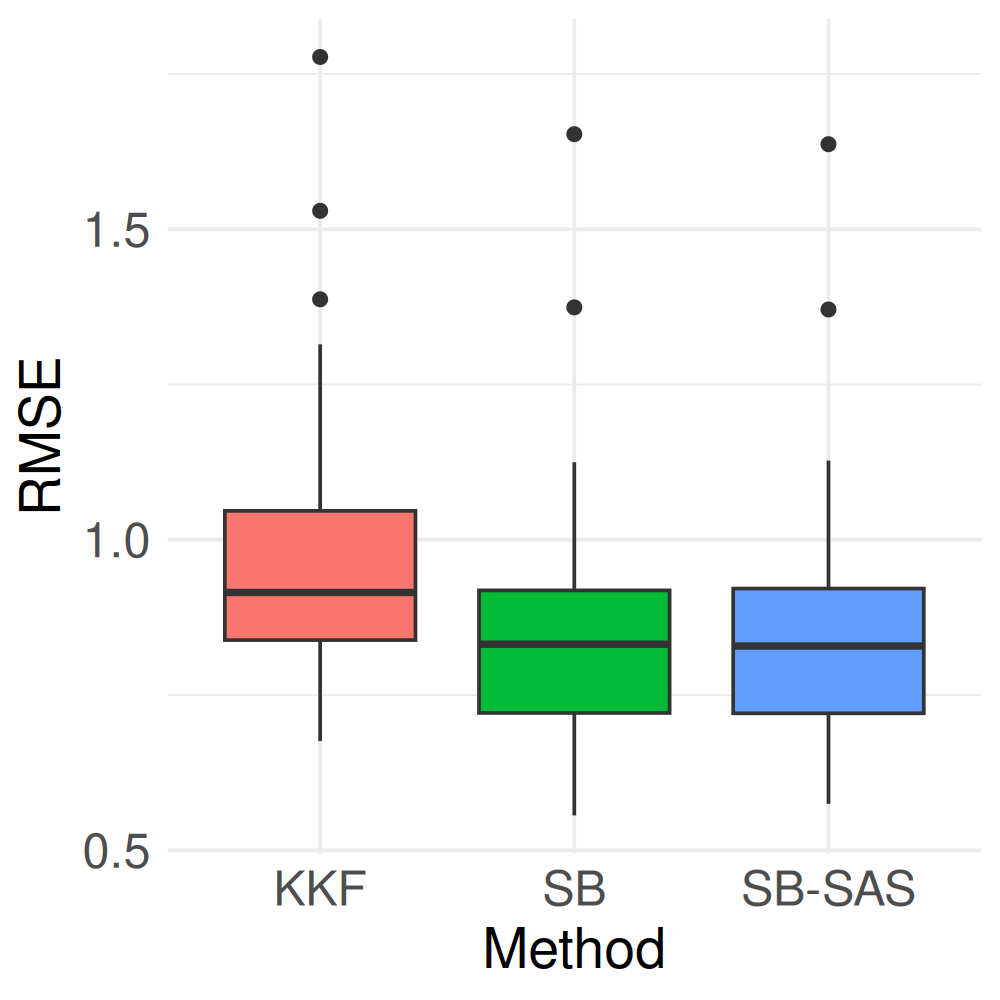}
        \caption*{$m=0.1$ \& $n=1$}
    \end{subfigure}
    \caption{Out-of-sample root mean squared error in predicting the observations at $5^\circ \mathrm{C}$ generated with different kinetics dynamics.}
    \label{fig:sim_vs}
\end{figure}

Next, we examine the predictive performance of different models.  Figure~\ref{fig:sim_vs} reports out-of-sample RMSE for predicting held-out observations at $5^\circ\mathrm{C}$. When the true dynamics are linear, $(m,n)=(0,0)$, KKF, SB, and SB-SAS have comparable prediction error, showing that the more flexible SB formulation does not lose much efficiency when the simpler KKF model is correct. In the nonlinear scenarios, $(m,n)=(0,1)$ and $(m,n)=(0.1,1)$, both SB-based methods improve over KKF. This suggests the SB-based methods can well approximate linear dynamics and generalize to nonlinear data generation mechanisms. 

Table~\ref{tab:vs_coverage} reports frequentist coverage and average length of $95\%$ posterior predictive intervals for held-out CQA values. In the linear scenario, all three methods achieve coverage close to the nominal level with similar interval lengths. Under nonlinear dynamics, KKF remains conservative, with coverage above the nominal level, but this comes at the cost of substantially wider intervals. This is likely caused by KKF model compensating its model misspecification and inability to capture the true dynamics with a larger estimate of residual error variance. By contrast, SB and SB-SAS maintain approximately nominal coverage with much shorter intervals. Thus, the improved prediction from the SB formulation is not obtained by inflating posterior uncertainty; rather, it comes from a more appropriate kinetic specification. The similarity between SB and SB-SAS also indicates that the variable-selection prior improves interpretability without materially compromising predictive uncertainty quantification.

\begin{table}[ht]
\centering
\resizebox{\textwidth}{!}{%
\begin{tabular}{lcc|cc|cc}
\toprule
& \multicolumn{2}{c|}{KKF}
& \multicolumn{2}{c|}{SB}
& \multicolumn{2}{c}{SB-SAS} \\
\cmidrule(lr){2-3} \cmidrule(lr){4-5} \cmidrule(lr){6-7}
Scenario
& Coverage & Length
& Coverage & Length
& Coverage & Length \\
\midrule
$m = 0,\ n = 0$
& $0.932$ $(0.073)$ & $7.788$ $(0.434)$
& $0.930$ $(0.072)$ & $7.814$ $(0.428)$
& $0.930$ $(0.072)$ & $7.706$ $(0.442)$ \\

$m = 0,\ n = 1$
& $0.988$ $(0.034)$ & $14.267$ $(0.656)$
& $0.945$ $(0.064)$ & $8.028$ $(0.431)$
& $0.943$ $(0.059)$ & $7.994$ $(0.430)$ \\

$m = 0.1,\ n = 1$
& $0.977$ $(0.041)$ & $10.743$ $(0.588)$
& $0.948$ $(0.063)$ & $7.962$ $(0.420)$
& $0.945$ $(0.062)$ & $7.936$ $(0.453)$ \\
\bottomrule
\end{tabular}%
}
\caption{Mean (standard deviation) of empirical coverage and credible interval length of KKF, SB, and SB-SAS models across the three variable-selection simulation scenarios.}
\label{tab:vs_coverage}
\end{table}

\subsection{Arrhenius vs non-Arrhenius behavior}\label{subsec:numerical_experiments_gae}

The second set of simulations evaluates robustness to misspecification of the temperature--rate relationship. We considered two data-generating mechanisms with the same observation model, time-temperature design, and Sestak--Berggren kinetic dynamics described above. In the Arrhenius scenario, the reaction rates were generated exactly from the classical Arrhenius equation,
\begin{equation*}
k(T)=k_{\text{ref}}\exp\left\{\frac{\exp(11.5)}{8.314}\left(\frac{1}{298.15}-\frac{1}{T}\right)\right\},
\end{equation*}
so that models imposing Arrhenius dynamics are correctly specified. In the non-Arrhenius scenario, we introduced systematic deviations at the extreme temperatures by perturbing the log-rates: the log-rate at $5^\circ\mathrm{C}$ was shifted by a draw from $N(-2,0.2^2)$, the log-rate at $40^\circ\mathrm{C}$ was shifted by a draw from $N(2,0.2^2)$, and the intermediate-temperature rates were left unchanged. Thus, the non-Arrhenius scenario preserves the same degradation mechanism but violates the log-linear Arrhenius relationship between inverse temperature and reaction rates. These two scenarios allow us to compare the efficiency of Arrhenius-based models when the physical assumption is correct with the robustness of the generalized Arrhenius formulation when the assumption is violated.

\begin{figure}[ht]
    \centering
    \begin{subfigure}{0.48\textwidth}
        \centering
        \includegraphics[width=\linewidth]{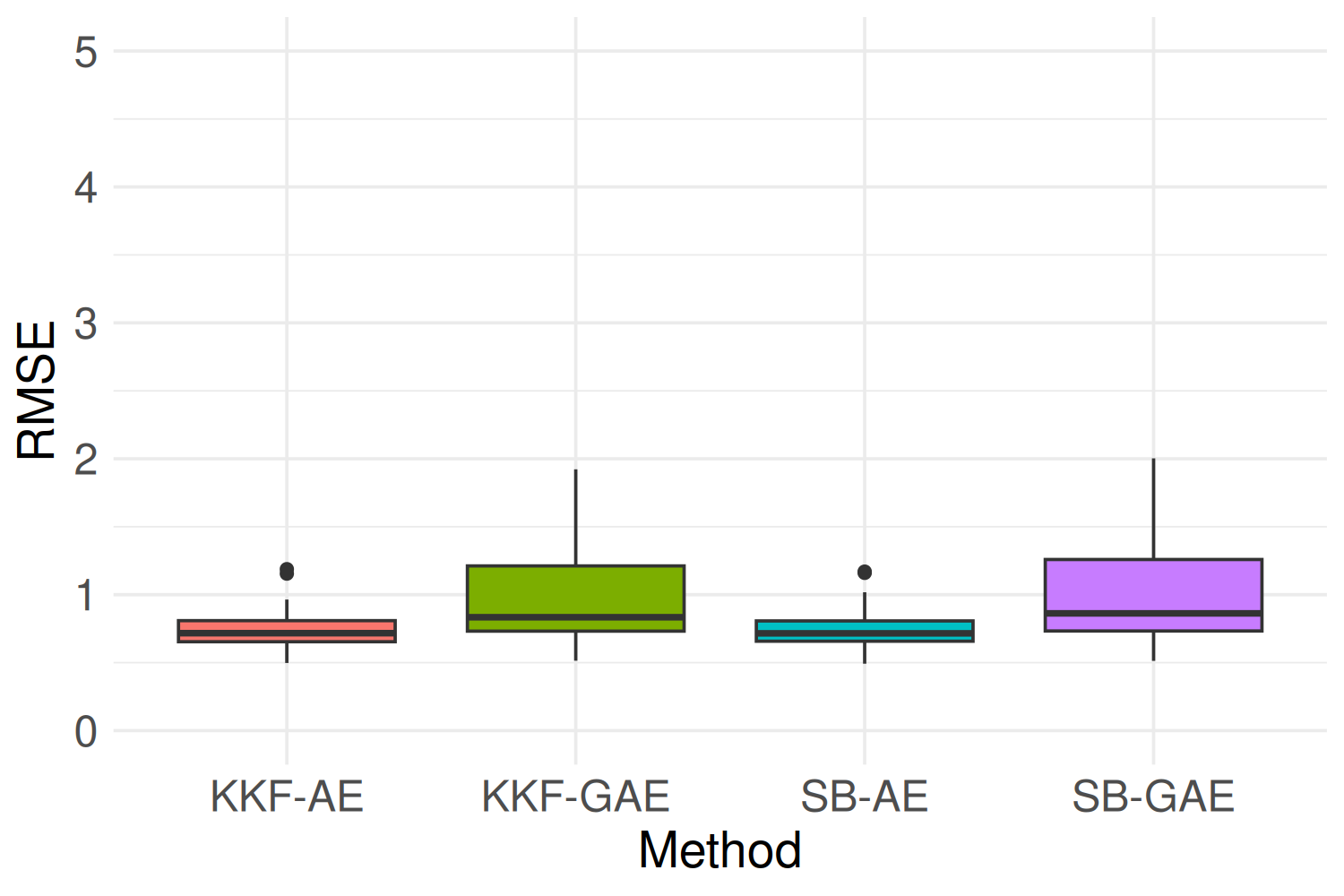}
        \caption*{$m=0, n=0$ \& Arrhenius dynamics}
    \end{subfigure}
    \hfill
    \begin{subfigure}{0.48\textwidth}
        \centering
        \includegraphics[width=\linewidth]{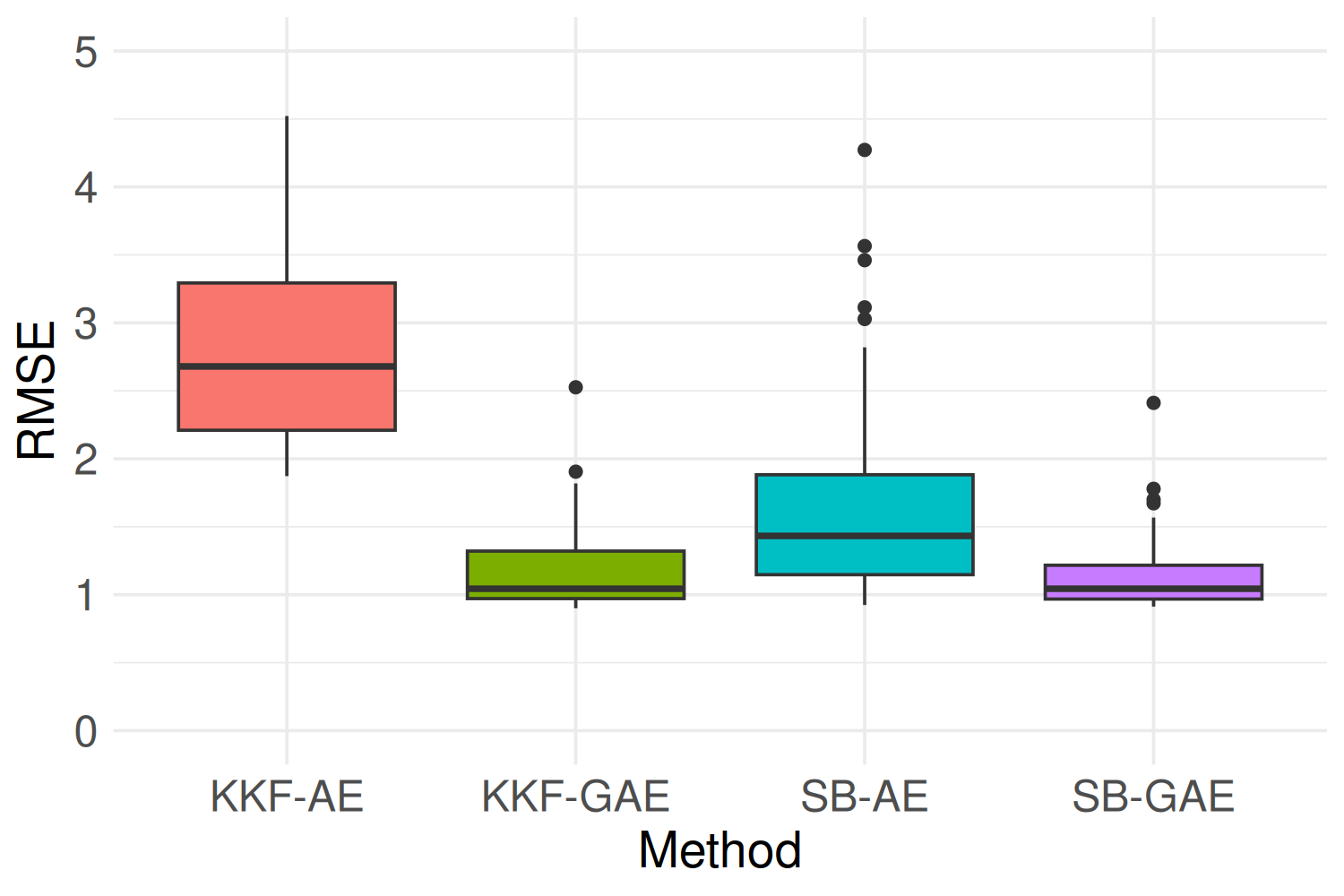}
        \caption*{$m=0, n=0$ \& generalized-Arrhenius dynamics}
    \end{subfigure}
    \vfill
    \begin{subfigure}{0.48\textwidth}
        \centering
        \includegraphics[width=\linewidth]{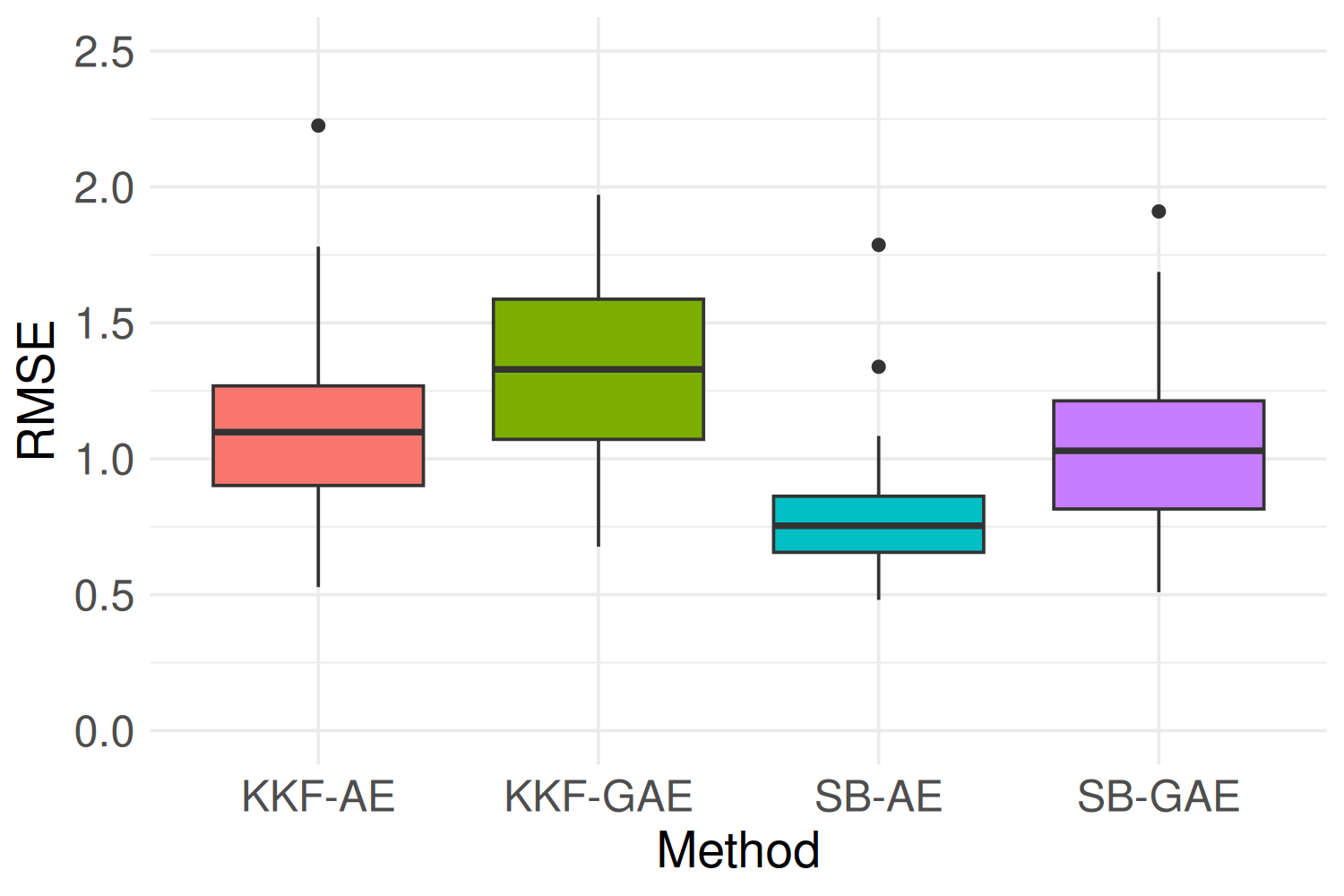}
        \caption*{$m=0, n=1$ \& Arrhenius dynamics}
    \end{subfigure}
    \hfill
    \begin{subfigure}{0.48\textwidth}
        \centering
        \includegraphics[width=\linewidth]{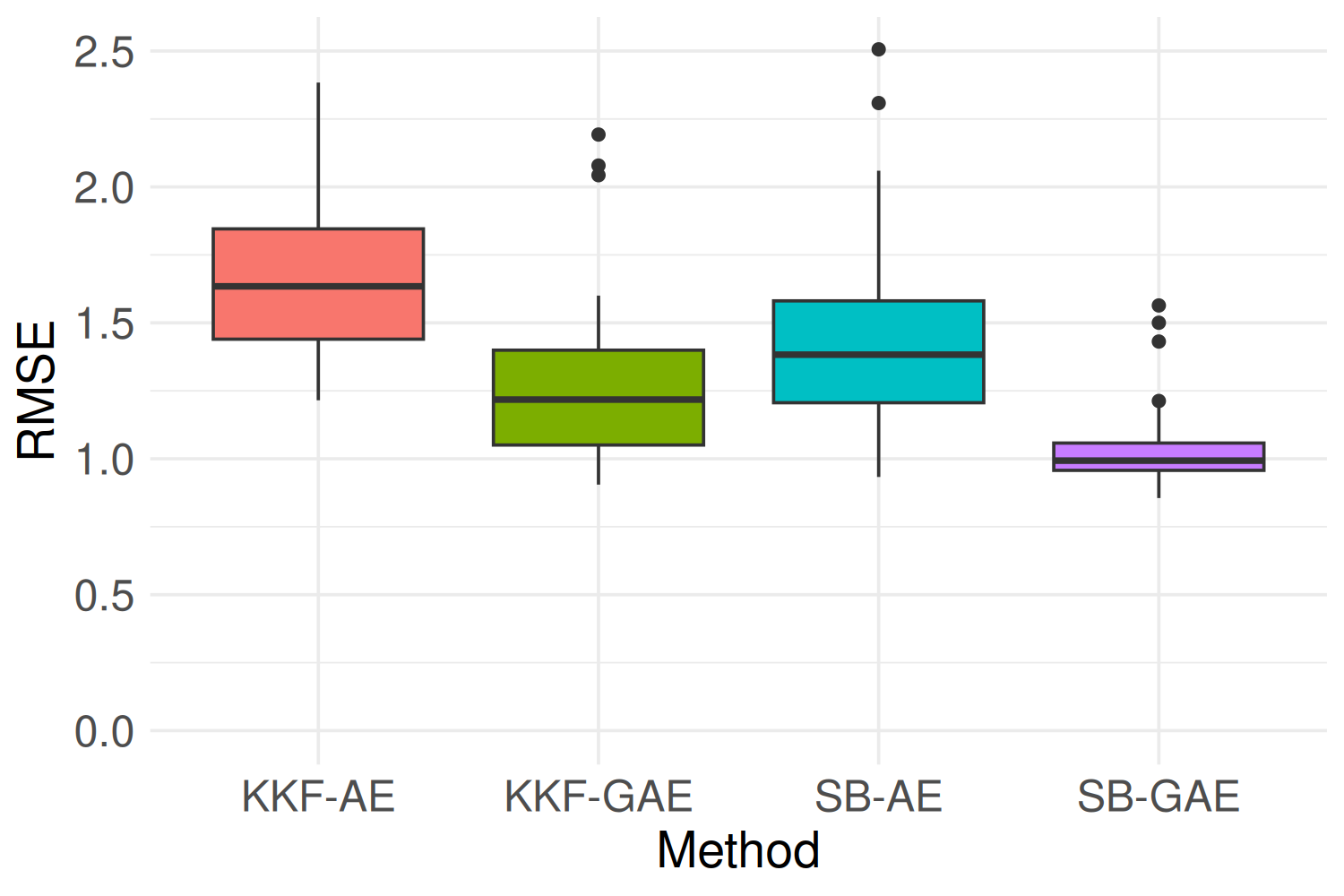}
        \caption*{$m=0, n=1$ \& generalized-Arrhenius dynamics}
    \end{subfigure}
    \caption{Out-of-sample root mean squared error in predicting the observations at $5^\circ \mathrm{C}$ generated with linear (top) and nonlinear (bottom) Arrhenius (left) and generalized-Arrhenius (right) dynamics.}
    \label{fig:sim_arr}
\end{figure}

Figure~\ref{fig:sim_arr} reports the out-of-sample root mean squared error in predicting the observations at $5^\circ \mathrm{C}$ under different scenarios. In particular, when observations follow Arrhenius dynamics (left panels), correctly specified models tend to have better performance but with a relatively small advantage compared to the GAE-formulations. In particular, AE-models impose the correct temperature-rate structure and estimate fewer parameters, determining an efficiency gain compared to the GAE counterpart. Nevertheless, the GAE formulations tend to pay only a moderate price, due to the prior formulation promoting shrinkage towards the Arrhenius dynamics. 
When Arrhenius relationship is misspecified (right panels), the opposite pattern arises. AE-based models suffer a large increase in prediction error, especially in the linear kinetics scenario (top panel), because the imposed log-linear temperature-rate relationship cannot accommodate the perturbed reaction rates at the extreme temperatures. Conversely, GAE formulations' performance remains robust and are substantial improvements over the misspecified counterparts. 

Table~\ref{tab:arr_coverage} reports coverage and interval length for AE and GAE models under Arrhenius and non-Arrhenius data-generating mechanisms. When Arrhenius is correctly specified, AE-based models are more efficient, producing shorter intervals than GAE-based models. This is the expected price of allowing deviations from Arrhenius when no such deviations are present. However, when the Arrhenius equation is violated, AE-based models become highly conservative and produce very wide intervals. For example, in the non-Arrhenius linear scenario, AE intervals have coverage equal to $>95\%$, whereas GAE achieves close-to-nominal coverage with intervals roughly three times shorter. A similar pattern appears in the nonlinear scenario: SB-GAE provides much shorter intervals than SB-AE while retaining approximately valid coverage. These results show that the GAE formulation pays a modest efficiency cost under correct Arrhenius specification but provides substantial robustness when the temperature--rate relationship is misspecified.

\begin{table}[ht]
\centering
\resizebox{\textwidth}{!}{%
\begin{tabular}{llcc|cc|cc|cc}
\toprule
& & \multicolumn{2}{c|}{KKF-AE}
& \multicolumn{2}{c|}{KKF-GAE}
& \multicolumn{2}{c|}{SB-AE}
& \multicolumn{2}{c}{SB-GAE} \\
\cmidrule(lr){3-4} \cmidrule(lr){5-6}
\cmidrule(lr){7-8} \cmidrule(lr){9-10}
Arrhenius? & Kinetic parameters
& Coverage & Length
& Coverage & Length
& Coverage & Length
& Coverage & Length \\
\midrule
\cmark
& $m = 0,\ n = 0$
& $0.932$ $(0.073)$ & $7.781$ $(0.433)$
& $0.930$ $(0.104)$ & $10.472$ $(0.680)$
& $0.935$ $(0.072)$ & $7.814$ $(0.427)$
& $0.932$ $(0.111)$ & $10.783$ $(0.779)$ \\

\cmark
& $m = 0,\ n = 1$
& $0.988$ $(0.034)$ & $14.256$ $(0.663)$
& $0.977$ $(0.063)$ & $14.578$ $(1.123)$
& $0.943$ $(0.062)$ & $8.030$ $(0.428)$
& $0.928$ $(0.115)$ & $10.687$ $(0.661)$ \\

\xmark
& $m = 0,\ n = 0$
& $0.983$ $(0.118)$ & $32.802$ $(6.439)$
& $0.957$ $(0.072)$ & $9.052$ $(0.974)$
& $0.983$ $(0.118)$ & $32.114$ $(7.039)$
& $0.965$ $(0.059)$ & $9.052$ $(1.026)$ \\

\xmark
& $m = 0,\ n = 1$
& $1.000$ $(0.000)$ & $28.277$ $(1.813)$
& $1.000$ $(0.000)$ & $21.668$ $(2.674)$
& $1.000$ $(0.000)$ & $24.024$ $(7.978)$
& $0.968$ $(0.051)$ & $8.865$ $(0.798)$ \\
\bottomrule
\end{tabular}%
}
\caption{Mean (standard deviation) of coverage and credible interval length under KKF-AE, KKF-GAE, SB-AE, and SB-GAE across the Arrhenius and generalized simulation scenarios.}
\label{tab:arr_coverage}
\end{table}

%\newpage

\section{Data analysis}\label{eq:applications}

\subsection{Non-linear dynamics and model selection for the \citet{huelsmeyer2023universal}'s data}

We first investigate whether accounting for non-linear degradation dynamics is necessary for reliable prediction, and whether the proposed framework can correctly adapt model complexity to the data, using the data from \citet{huelsmeyer2023universal}. We consider the proportion of acidic variants as the CQA of interest. The dataset contains 53 observations collected across 6 temperatures ($5^\circ \mathrm{C}$, $25^\circ \mathrm{C}$, $30^\circ \mathrm{C}$, $35^\circ \mathrm{C}$, $40^\circ \mathrm{C}$, $45^\circ \mathrm{C}$), which are summarized in Table \ref{tab:app_1}. Our primary interest is predicting the long-term evolution of this attribute at $5^\circ \mathrm{C}$, the intended refrigerated storage temperature.
\begin{table}[ht]
\centering
\begin{tabular}{c|ccccccccccccccc}
\hline
& \multicolumn{15}{c}{\textbf{Time (days)}}\\
\textbf{Temperature} 
& \textbf{0} & \textbf{14} & \textbf{28} & \textbf{42} & \textbf{56} 
& \textbf{70} & \textbf{84} & \textbf{91} & \textbf{98} 
& \textbf{182} & \textbf{273} & \textbf{365} 
& \textbf{547} & \textbf{730} & \textbf{1095} \\
\hline
$5^\circ\mathrm{C}$  
& 2 & 0 & 0 & 0 & 0 & 0 & 0 & 2 & 0 & 2 & 2 & 2 & 1 & 1 & 1 \\

$25^\circ\mathrm{C}$ 
& 1 & 1 & 1 & 1 & 1 & 1 & 1 & 0 & 1 & 0 & 0 & 0 & 0 & 0 & 0 \\

$30^\circ\mathrm{C}$ 
& 1 & 1 & 1 & 1 & 1 & 1 & 1 & 0 & 1 & 0 & 0 & 0 & 0 & 0 & 0 \\

$35^\circ\mathrm{C}$ 
& 1 & 1 & 1 & 1 & 1 & 1 & 1 & 0 & 1 & 0 & 0 & 0 & 0 & 0 & 0 \\

$40^\circ\mathrm{C}$ 
& 1 & 1 & 1 & 1 & 1 & 1 & 1 & 0 & 1 & 0 & 0 & 0 & 0 & 0 & 0 \\

$45^\circ\mathrm{C}$ 
& 1 & 1 & 1 & 1 & 1 & 1 & 1 & 0 & 1 & 0 & 0 & 0 & 0 & 0 & 0 \\
\hline
\end{tabular}
\caption{Counts of samples by temperature and time from storage}\label{tab:app_1}
\end{table}

\begin{figure}
    \centering
        \includegraphics[width=\linewidth]{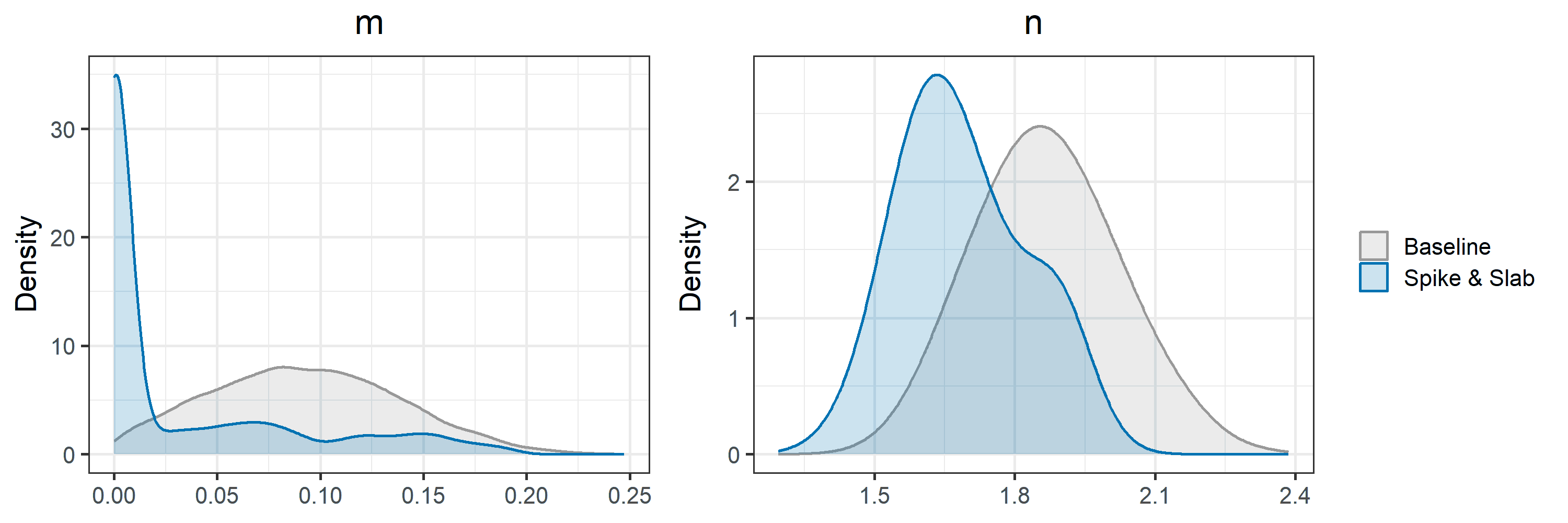}
    \caption{Posterior distributions of the autocatalytic parameter $m$ (left) and the reaction order $n$ (right) for the SB model with Gaussian prior (grey) and the SB model with spike and slab priors (light blue). }
    \label{fig:1_variable_selection}
\end{figure}

We compare the standard KKF model, which assumes linear degradation dynamics, with the SB model that allows for non-linear kinetics. Models are fit using observations within the first 6 months and evaluated on held-out data to assess predictive performance. 
\begin{figure}
    \centering
        \includegraphics[width=\linewidth]{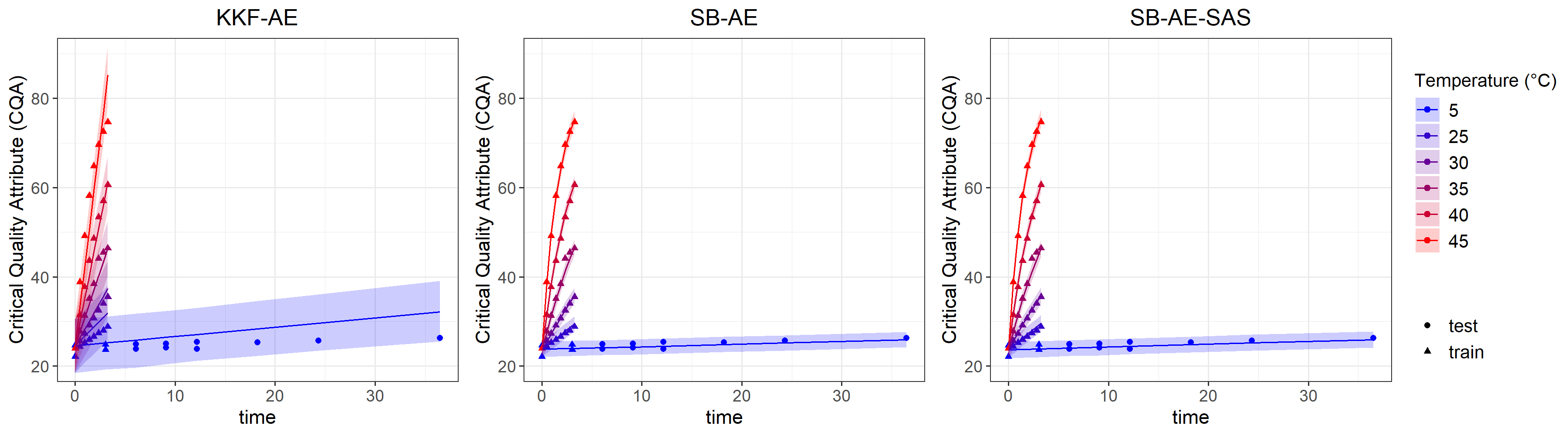}
    %\caption{$n=500$} 
    \caption{In-sample fit and out-of-sample predictions for KKF model (left panel), SB model (middle panel), and SB model with spike and slab (SAS) priors (right panel). AE denotes models assuming the Arrhenius equation.}
    \label{fig:1_predictions}
\end{figure}
Figure \ref{fig:1_predictions} summarizes in-sample fits and out-of-sample predictions. The KKF model (left panel) is clearly misspecified, producing severe overestimation of the CQA at  $5^\circ \mathrm{C}$  over time. This is problematic from a decision-making perspective, as it would incorrectly suggest instability at storage conditions and could lead to discarding a viable compound.
In contrast, the SB model (middle panel) captures the underlying non-linear dynamics and yields substantially improved predictions. The SB model with spike-and-slab priors (right panel) performs comparably, indicating that adaptive model selection does not degrade predictive accuracy. 
The data favor nonlinear but non-autocatalytic dynamics, indicating that higher-order effects are present while autocatalytic behavior is not supported. Indeed, posterior inference suggests no strong evidence for the inclusion of the autocatalytic parameter $m$, with its slab probability concentrated near zero. 
In contrast, the reaction order parameter $n$ is clearly supported by the data, with posterior mean $1.68$ and $95\%$ credible interval $(1.46, 1.97)$, closely matching the estimates from the SB model with truncated Gaussian priors (Figure \ref{fig:1_variable_selection}).

\begin{figure}
    \centering
        \includegraphics[width=\linewidth]{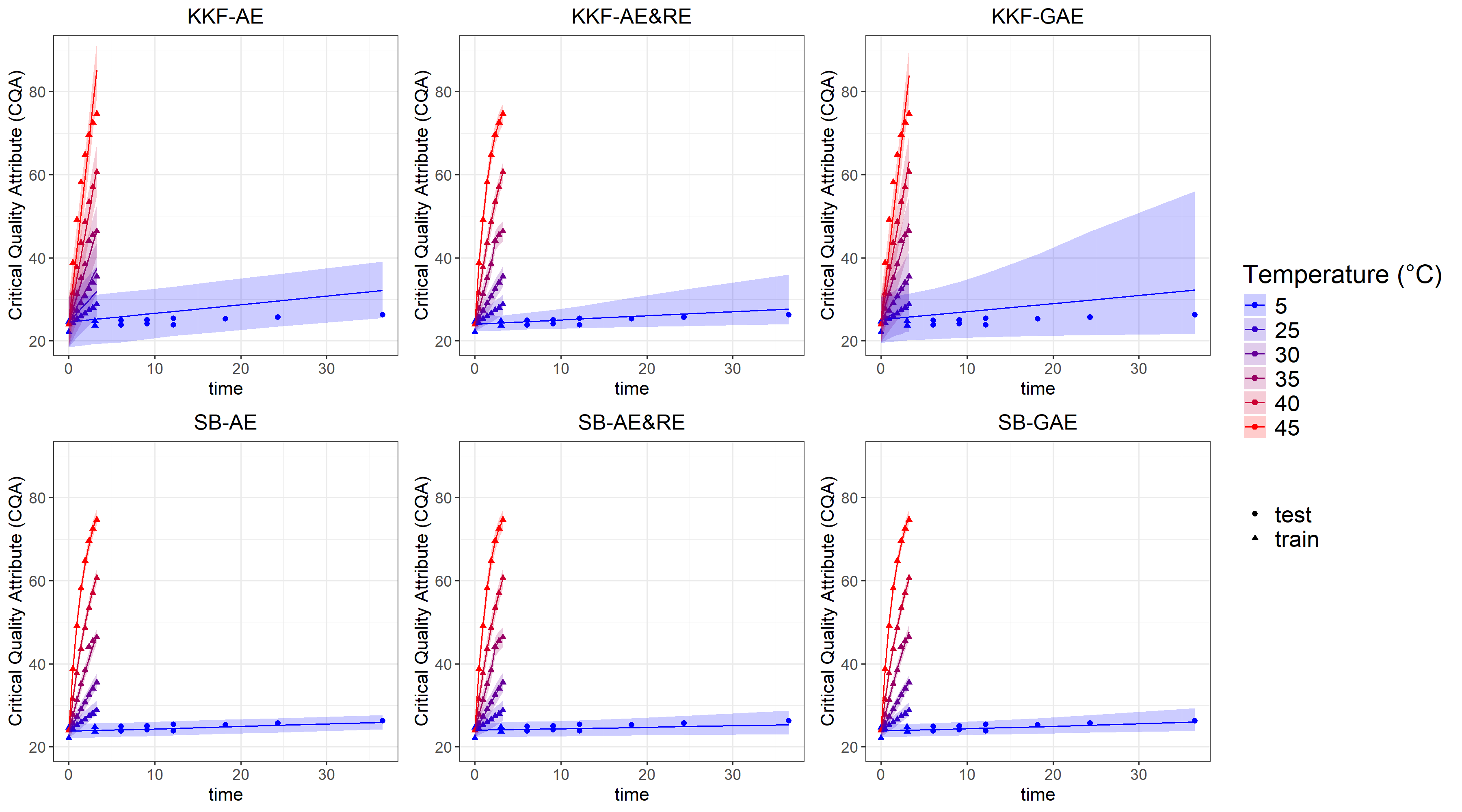}
    %\caption{$n=500$} 
    \caption{In-sample fit and out-of-sample predictions for KKF model (top panels), SB model (bottom panels) with the Arrhenius equation (AE) (left panels), with the Arrhenius equation with random effects (AE\&RE) (middle panels), and the generalized Arrhenius equation framework (GAE) (right panels).}
    \label{fig:1_predictions_ae}
\end{figure}

Next, we also examined sensitivity to the Arrhenius assumption, by fitting KKF and SB models under the generalized Arrhenius (GAE) framework described in Section \ref{subsec:generalized_Arrhenius}, as well as models incorporating multiplicative random effects (AE\&RE) \citep{chau2023} (Figure \ref{fig:1_predictions_ae}). Adopting the GAE framework leaves point estimates largely unchanged, while increasing predictive uncertainty, particularly for the KKF model, reflecting the additional model flexibility. This indicates that the Arrhenius-informed prior shrinks toward classical Arrhenius behavior when appropriate, avoiding unnecessary complexity.
In contrast, in this case, introducing random effects partially compensates for model misspecification in KKF, while still producing larger prediction intervals than the corresponding SB model. However, as we show in the next example, such an approach can be unstable in more complex scenarios.

\pagebreak
\subsection{Assessing Arrhenius behaviour of \citet{clenet}'s data}\label{subsec:application_gae}
We now focus on a central challenge in predictive stability: robustness of extrapolation under violations of Arrhenius behavior.
We analyze the data from \citet{clenet}, with antigenicity assessed by enzyme-linked immunosorbent assay (ELISA) as the CQA of interest. 
The dataset (summarized in Table \ref{tab:app_2}) contains 96 samples recorded at 4 temperatures ($5^\circ \mathrm{C}$, $25^\circ \mathrm{C}$, $37^\circ \mathrm{C}$, $45^\circ \mathrm{C}$). We fit the models using the 93 samples recorded within $6$ months of storage. 
Our primary interest is long-term prediction at $5^\circ \mathrm{C}$, assessed using the three held-out observations collected after 1020 days of storage.

\begin{table}[ht]
\centering
\begin{tabular}{c|cccccccccc}
\hline
& \multicolumn{10}{c}{\textbf{Time (days)}}\\
\textbf{Temperature} & \textbf{0} & \textbf{1} & \textbf{3} & \textbf{7} & \textbf{13} & \textbf{30} & \textbf{60} & \textbf{90} & \textbf{180} & \textbf{1020} \\
\hline
$5^\circ\mathrm{C}$  & 6 & 0 & 0 & 0 & 0 & 3 & 0 & 3 & 3 & 3 \\
$25^\circ\mathrm{C}$ & 6 & 0 & 0 & 3 & 3 & 3 & 3 & 3 & 3 & 0 \\
$37^\circ\mathrm{C}$ & 6 & 3 & 3 & 3 & 3 & 3 & 3 & 3 & 3 & 0 \\
$45^\circ\mathrm{C}$ & 6 & 3 & 3 & 3 & 3 & 3 & 0 & 3 & 0 & 0 \\
\hline
\end{tabular}
\caption{Counts of samples by temperature and time from storage}\label{tab:app_2}
\end{table}

\begin{figure}
    \centering
        \includegraphics[width=\linewidth]{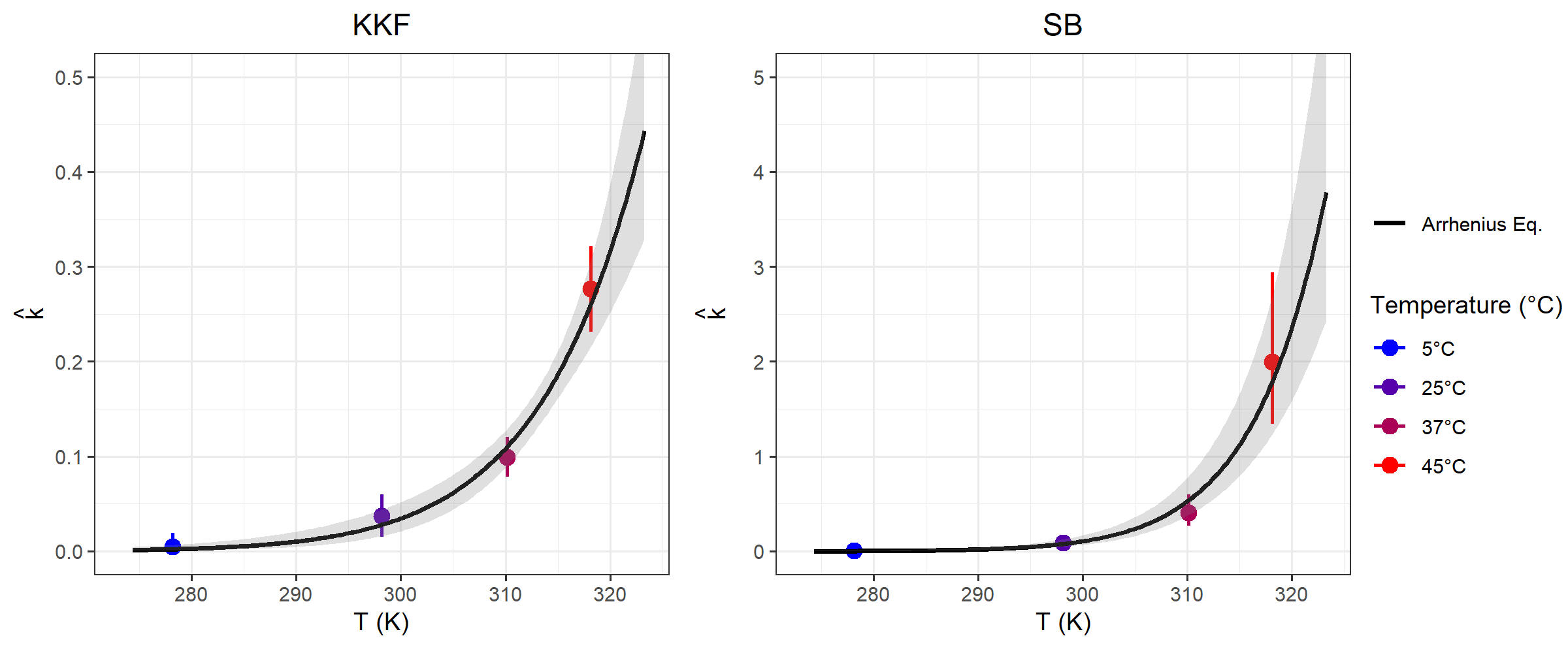}
        \hfill
         \includegraphics[width=\linewidth]{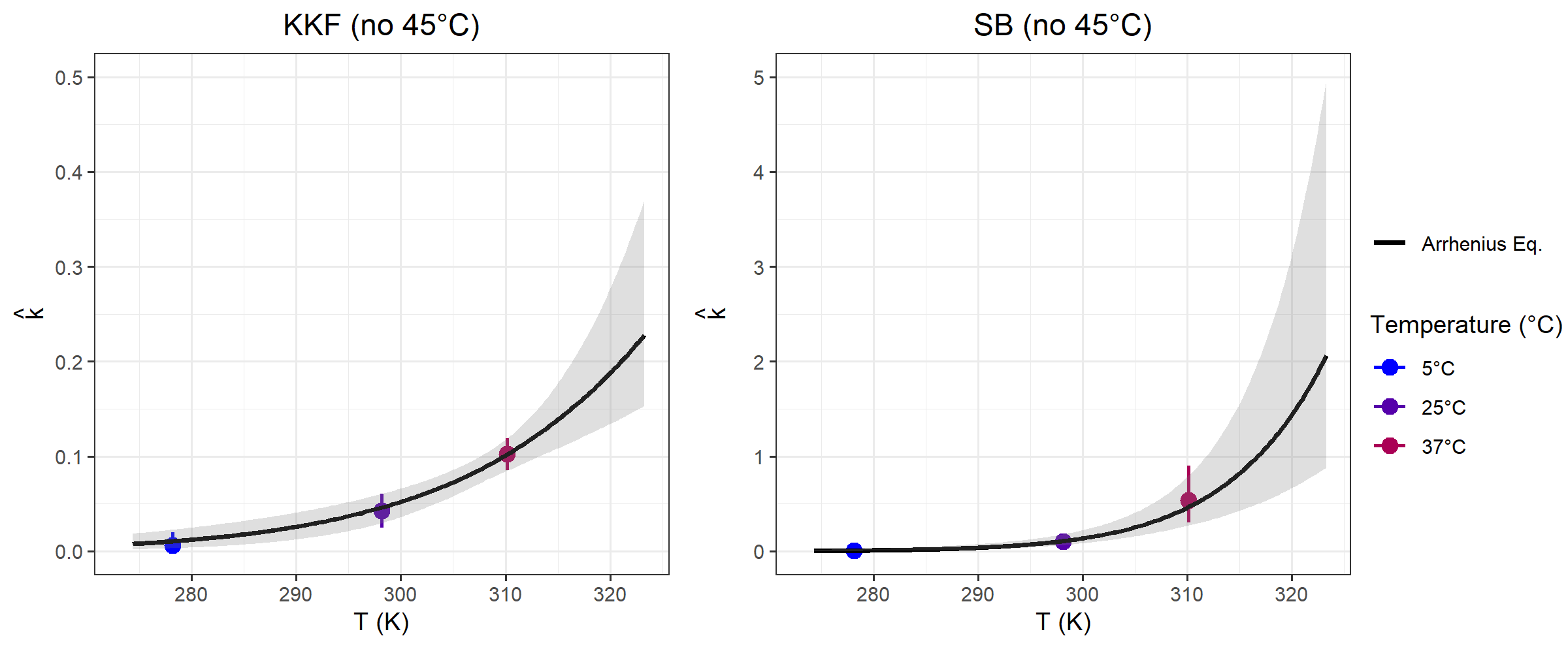}
    \caption{Arrhenius plot (temperature on x-axis vs reaction rate on the y-axis) for the KKF model (left panel) and SB model (right panel) with the complete dataset (top panels) and removing the $45^\circ \mathrm{C}$ data (bottom panels). The black solid line (shaded area) shows the estimated relationship (and its $95\%$ confidence band) between temperature and reaction rates with models assuming the Arrhenius equation, dots (vertical lines) show estimates ($95\%$ credible intervals) under the generalized Arrhenius equation framework. }
    \label{fig:8_Arrhenius}
\end{figure}

A key practical question is whether to include data from temperatures higher than $40^\circ \mathrm{C}$, where Arrhenius assumptions can be violated, Such conditions can induce non-Arrhenius kinetics \citep{waterman_adami}. We investigate the potential violations of the Arrhenius equation in Figure \ref{fig:8_Arrhenius}. The solid lines show the estimated relationship between the temperatures and the reaction rates from the KKF (left) and SB (right) models using the Arrhenius equations (equation \eqref{eq:Arrhenius_eq}), while dots represent estimates for the reaction rates under our GAE framework.
 The data shows non-negligible evidence of violation from the Arrhenius dynamics for samples stored at $45^\circ \mathrm{C}$. In particular, there is evidence that the reaction rate at such temperature might be higher than the one prescribed by equation \eqref{eq:Arrhenius_eq}. 

This raises a fundamental trade-off: including high-temperature data may introduce bias due to model misspecification, while excluding it reduces sample size and increases estimation variability. 
This issue is particularly problematic when models fitted without such samples produce estimates that are completely different than the ones from models fitted on the complete dataset. In these cases, it is often hard to understand whether the differences arise from the deviation from the Arrhenius behaviour of samples at high temperature or from the increased instability of estimates when such measurements are discarded, due to the lower sample-size. 

\begin{figure}
    \centering
        \includegraphics[width=\linewidth]{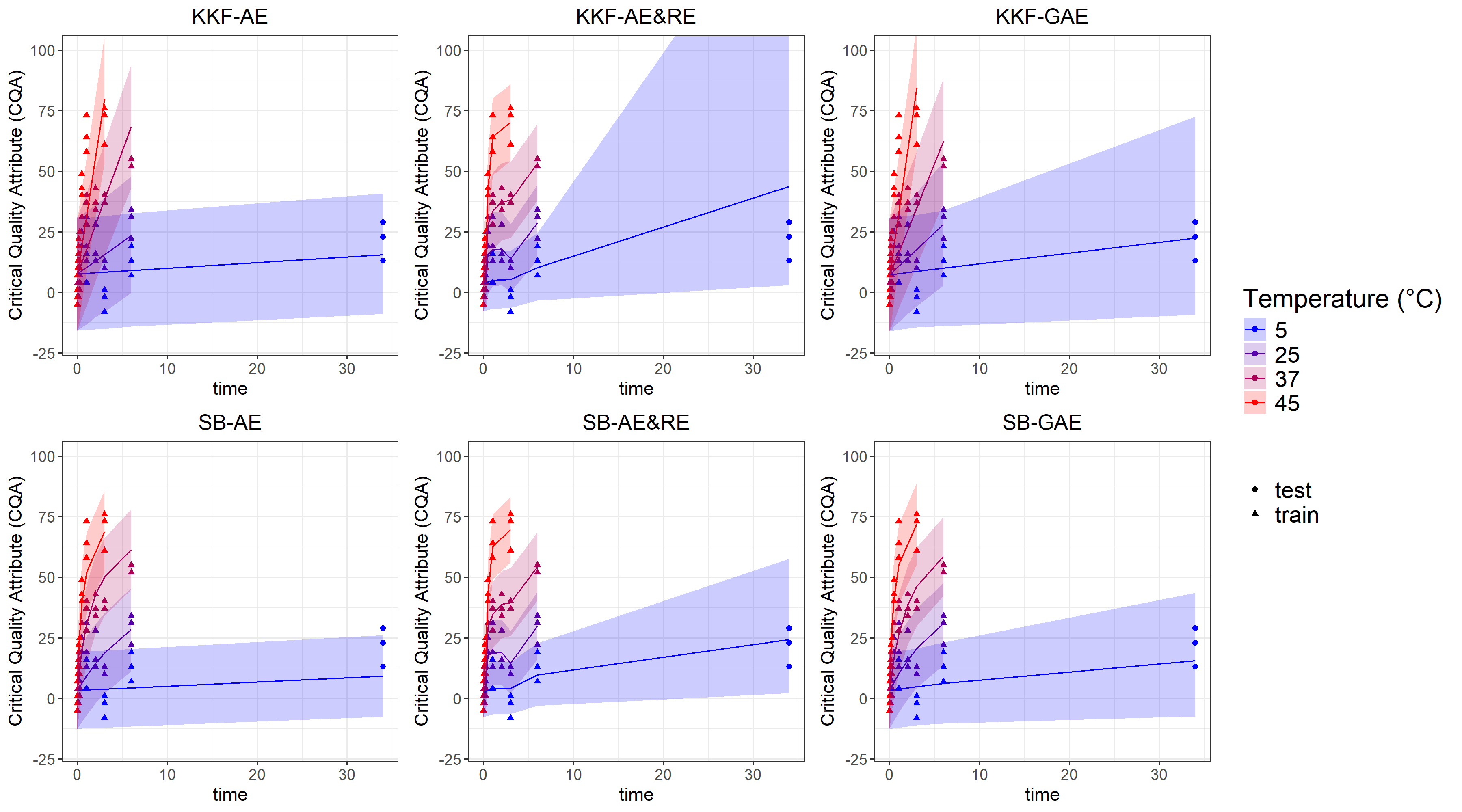}
    \caption{In-sample fit and out-of-sample predictions for KKF model (top panels) and SB model (bottom panels), 
    with the Arrhenius equation without (AE) and with random effect (AE\&RE) (left and middle panels), and the generalized Arrhenius equation framework (GAE) (right panels).}
    \label{fig:8_predictions_all}
\end{figure}

\begin{figure}
    \centering
        \includegraphics[width=\linewidth]{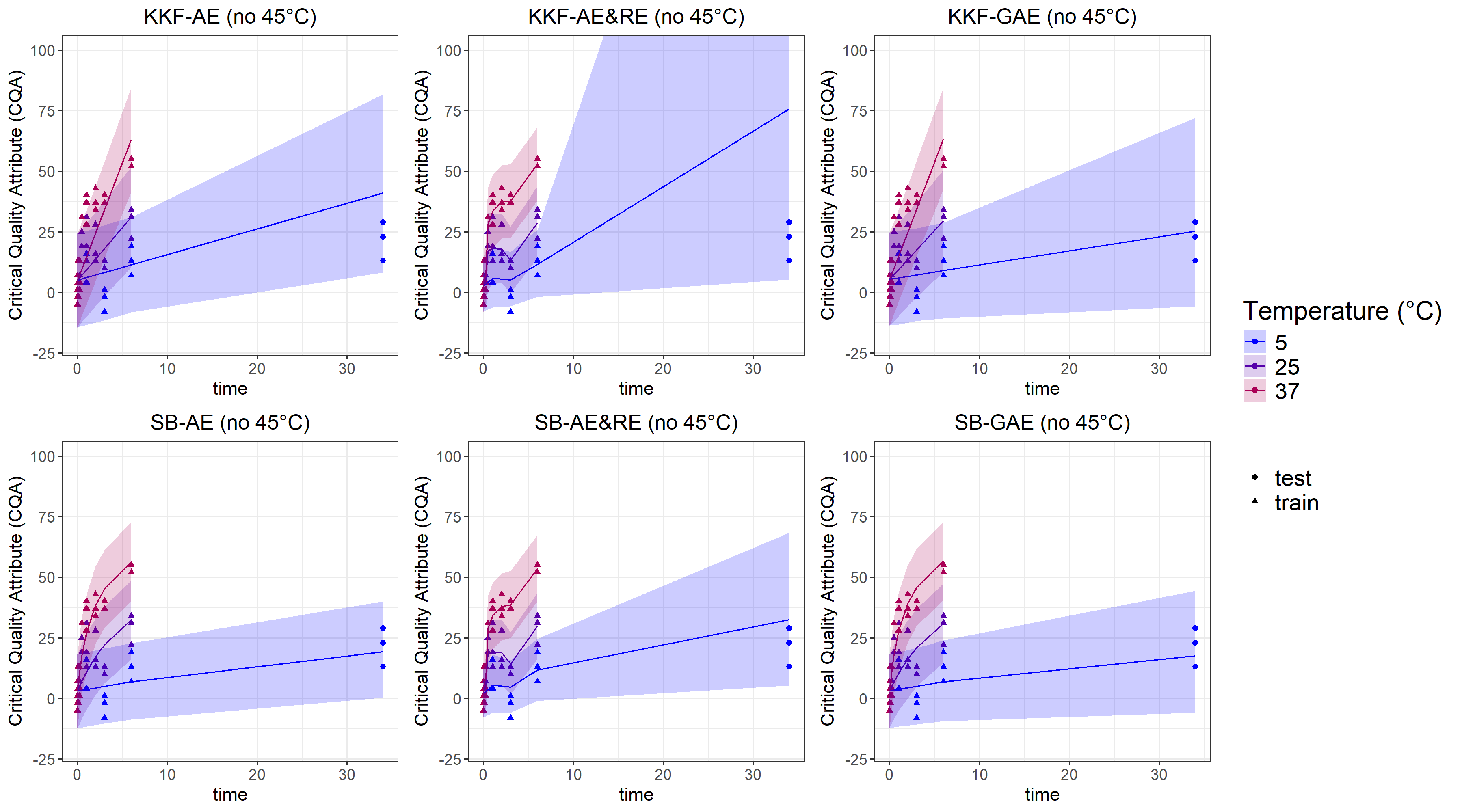}
    \caption{In-sample fit and out-of-sample predictions for KKF model (top panels) and SB model (bottom panels) fitted without the $45^\circ \mathrm{C}$ data, 
    with the Arrhenius equation without (AE) and with random effect (AE\&RE) (left and middle panels), and the generalized Arrhenius equation framework (GAE) (right panels).}
    \label{fig:8_predictions_no45}
\end{figure}

\begin{figure}
    \centering
        \includegraphics[width=\linewidth]{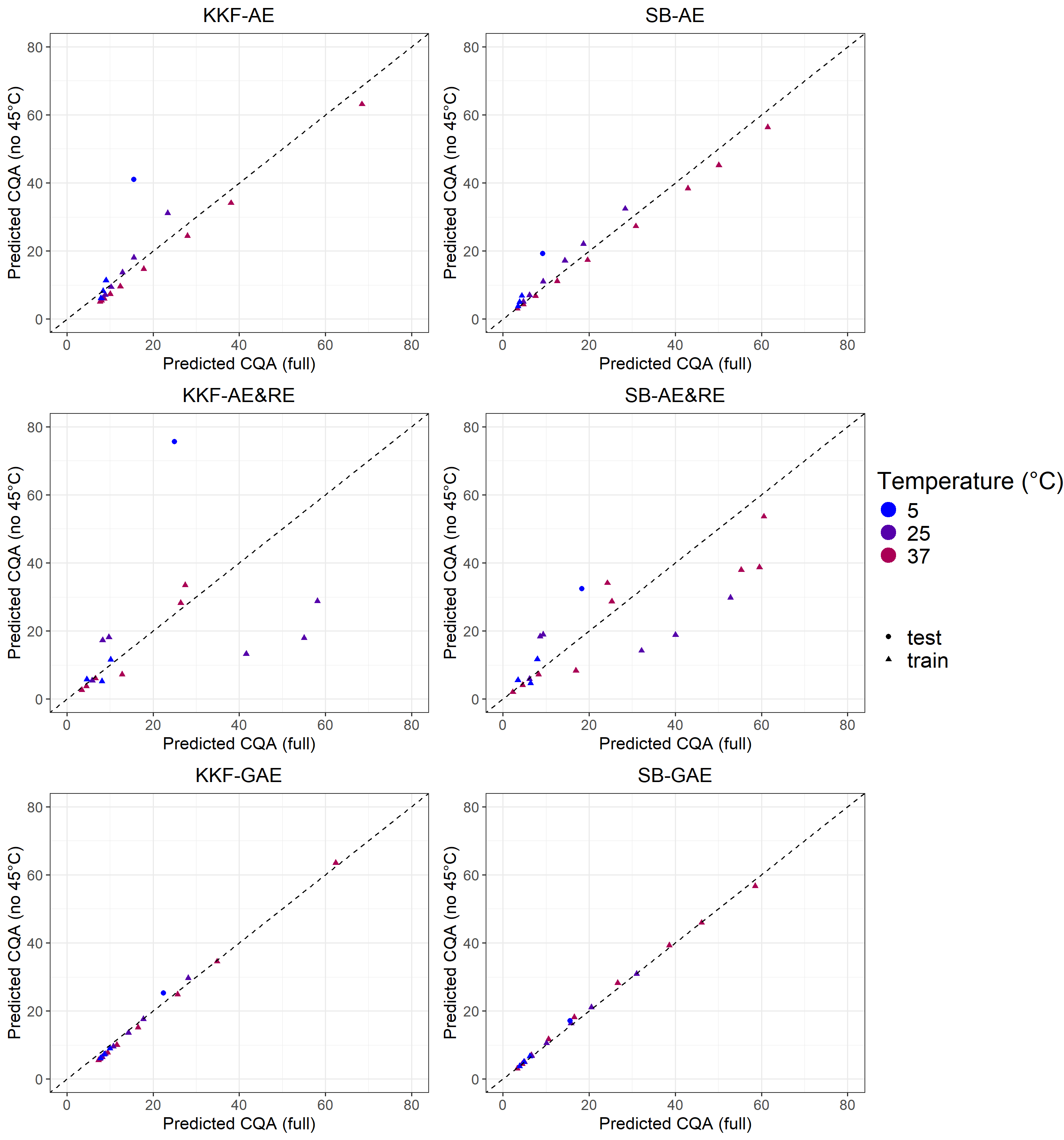}
    \caption{Sensitivity of fitted values and predictions to the inclusion of the $45^\circ \mathrm{C}$ data for KKF model (LEFT panels) and SB model (RIGHT panels) with the Arrhenius equation without (AE) and with random effect (AE\&RE) (top and middle panels), and the generalized Arrhenius equation framework (GAE) (bottom panels). Predicted CQA from models fitted with the entire dataset (x-axis) and removing the $45^\circ \mathrm{C}$ data (y-axis)}
    \label{fig:8_sensitivity_45}
\end{figure}

To assess robustness, we perform a stress test by removing the 45°C data and comparing predictions to those obtained using the full dataset (Figures \ref{fig:8_predictions_all}–\ref{fig:8_sensitivity_45}).
Figures \ref{fig:8_predictions_all} and \ref{fig:8_predictions_no45} show model fits and predictions using the complete datasets and discarding the $45^\circ \mathrm{C}$ data, respectively, and Figure \ref{fig:8_sensitivity_45} compares fitted and predicted values under the two scenarios. 
The generalized Arrhenius equation framework provides the most reliable and robust performance across all settings. GAE-based models consistently achieve accurate in-sample fits and out-of-sample predictions, while remaining stable when the $45^\circ \mathrm{C}$ data are removed. This robustness is particularly important, as it indicates that the model is not overly reliant on extreme temperature regimes to anchor its predictions. Instead, the GAE framework is sufficiently flexible to capture deviations from classical Arrhenius behavior leading to more credible extrapolation across time.

In contrast, the standard Arrhenius-based model with linear dynamics appears to perform adequately when fit on the full dataset, but this behavior is misleading. Once the $45^\circ \mathrm{C}$ data are excluded, their predictions change substantially, revealing a strong dependence on high-temperature observations.
This suggests that their apparent success is driven by a form of double misspecification, combining incorrect linear dynamics with violations of Arrhenius behavior, leading to unreliable extrapolation at lower temperatures.

Models incorporating random effects (AE\&RE) exhibit pronounced instability, with highly variable fits and poor predictive accuracy across both scenarios. This is likely due to the fact random effects absorb heterogeneity but do not fix structural misspecification in kinetics determining poor extrapolation results. 
This instability is further illustrated in Figure \ref{fig:8_sensitivity_45}: predictions from GAE-based models lie close to the identity line when comparing fits with and without $45^\circ \mathrm{C}$ data, while competing methods show substantial deviations, indicating sensitivity to data inclusion.

Overall, these results demonstrate that robust extrapolation in predictive stability requires explicitly modeling deviations from Arrhenius behavior; the GAE framework is the only approach that consistently achieves stable predictions.

\section{Conclusions}
We developed a Bayesian framework for predictive stability modeling that jointly accommodates nonlinear degradation dynamics and deviations from Arrhenius kinetics.

By leveraging the Sestak–Berggren formulation, the proposed framework allows for flexible modeling of degradation dynamics. The variable selection approach further supports interpretability and model development by identifying when simpler kinetic specifications are sufficient. Our results show that standard Arrhenius-based approaches can yield unstable and potentially misleading extrapolations, particularly when high-temperature data might deviated from the Arrhenius dynamics. In contrast, the proposed framework provides more stable predictions and more realistic uncertainty quantification across a range of scenarios. This robustness is critical in practice, where stability decisions must often be made based on limited and heterogeneous data.

From an applied perspective, the proposed approach offers a principled tool for model-informed shelf-life determination, enabling more reliable use of accelerated stability data while mitigating risks associated with model misspecification. More broadly, these results highlight the importance of incorporating flexible, data-adaptive structures within physically motivated models to support decision-making in pharmaceutical development. 

Several directions for future work are worth exploring. First, while the proposed framework accounts for deviations from Arrhenius kinetics, further extensions could address misspecification of the kinetic model itself, for example when the Sestak–Berggren formulation does not adequately capture the underlying degradation mechanism. Second, combining multiple modeling strategies, such as Arrhenius and non-Arrhenius formulations or alternative kinetic specifications (e.g., SB and KKF), through Bayesian model averaging \citep{hoeting1999bma} or related approaches could improve predictive performance, particularly when different models perform well in different temperature regimes. Finally, developing formal procedures to assess the validity of modeling assumptions, including testing for Arrhenius behavior or for specific kinetic structures (e.g., values of the reaction order and autocatalytic parameters), would provide additional guidance for model selection and increase the reliability of downstream stability decisions.
 
\section*{Acknowledgement}
LM was partially supported by the Office of Naval Research (Grant N00014-24-1-2626), and by the European Research Council under the European Union’s Horizon 2020 research and innovation programme (grant agreement No 856506).

\bibliographystyle{apalike}
\bibliography{references}

@article{sb_model,
  title={Combined kinetic analysis of solid-state reactions: a powerful tool for the simultaneous determination of kinetic parameters and the kinetic model without previous assumptions on the reaction mechanism},
  author={P{\'e}rez-Maqueda, LA and Criado, JM and Sanchez-Jimenez, PE},
  journal={The Journal of Physical Chemistry A},
  volume={110},
  number={45},
  pages={12456--12462},
  year={2006},
  publisher={ACS Publications}
}

@Misc{stan,
    title={{RStan}: the {R} interface to {Stan}},
    author={{Stan Development Team}},
    note={R package version 2.32.7},
    year={2025},
    url={https://mc-stan.org/},
}

@article{nuts,
  title={The {No-U-Turn} sampler: adaptively setting path lengths in {H}amiltonian {M}onte {C}arlo.},
  author={Hoffman, Matthew D and Gelman, Andrew and others},
  journal={J. Mach. Learn. Res.},
  volume={15},
  number={1},
  pages={1593--1623},
  year={2014}
}

@article{huelsmeyer2023universal,
  title={A universal tool for stability predictions of biotherapeutics, vaccines and in vitro diagnostic products},
  author={Huelsmeyer, M and Kuzman, Drago and Bon{\v{c}}ina, M and Martinez, J and Steinbrugger, C and Weusten, J and Calero-Rubio, C and Roche, W and Niederhaus, B and VanHaelst, Y and others},
  journal={Scientific reports},
  volume={13},
  number={1},
  pages={10077},
  year={2023},
  publisher={Nature Publishing Group UK London}
}

@article{bajaj2012,
title={Stability testing of pharmaceutical products},
author={Bajaj, Sanjay and Singla, Dinesh and Sakhuja, Neha},
journal={Journal of Applied Pharmaceutical Science},
volume={2},
number={3},
pages={129-138},
year={2012}
}

@article{kuzman2021,
  title = {Long-term stability predictions of therapeutic monoclonal antibodies in solution using Arrhenius-based kinetics},
  author = {Kuzman, D. and Bunc, M. and Ravnik, M. and Reiter, F. and {\v{Z}}agar, L. and Bon{\v{c}}ina, M.},
  journal = {Scientific Reports},
  volume = {11},
  pages = {20534},
  year = {2021}
}

@article{ferrari2025,
author = {Ferrari, Federico and Berger, Jordan and Lemieux, Linda and Paduraru, Crina and Dillon, Michael and Liaw, Andy and Carrillo, Ralf and Wong, Sally and Salami, Hossein and Avalle, Paolo and Sherer, Edward and Richardson, Douglas and Skomski, Daniel},
year = {2025},
month = {05},
pages = {17333},
title = {Bayesian hierarchical model predicts biopharmaceutical stability indicators and shelf life with application to multivalent human papillomavirus vaccine},
volume = {15},
journal = {Scientific Reports}
}

@article{kayser2011,
title={Evaluation of non-{A}rrhenius model for therapeutic monoclonal antibody aggregation},
  author={Veysel Kayser and Naresh Chennamsetty and Vladimir Voynov and Bernhard Helk and Kurt Forrer and Bernhardt L. Trout},
journal={Journal of Pharmaceutical Sciences},
volume={100},
number={7},
pages={2526-2542},
year={2011}
}

@article{wangroberts2013,
title={Non-{A}rrhenius protein aggregation},
  author={Wei Wang and Christopher J. Roberts},
journal={The AAPS Journal},
volume={15},
number={3},
pages={840-851},
year={2013}
}

@article{bunc2022,
  title   = {Aggregation time machine: A platform for the predictions and optimization of long-term antibody stability using short-term kinetic analyses},
  author  = {Bunc, M. and Had\v{z}i, S. and Graf, C. and Bon\v{c}ina, M. and Lah, J.},
  journal = {Journal of Medicinal Chemistry},
  volume  = {65},
  pages   = {2623--2632},
  year    = {2022}
}

@article{chau2023,
author = {Chau, Joris and Altan, Stan and Burggraeve, Anneleen and Coppenolle, Hans and Kifle, Yimer Wasihun and Prokopcova, Hana and Daele, Timothy and Sterckx, Hans},
year = {2023},
month = {11},
pages = {},
title = {A {B}ayesian Approach to Kinetic Modeling of Accelerated Stability Studies and Shelf Life Determination},
volume = {24},
journal = {AAPS PharmSciTech},
doi = {10.1208/s12249-023-02695-5}
}

@article{dormand_prince,
title = {A family of embedded Runge-Kutta formulae},
journal = {Journal of Computational and Applied Mathematics},
volume = {6},
number = {1},
pages = {19-26},
year = {1980},
issn = {0377-0427},
doi = {https://doi.org/10.1016/0771-050X(80)90013-3},
author = {J.R. Dormand and P.J. Prince}
}

@book{atkins2018physical,
  author    = {Atkins, Peter and de Paula, Julio and Keeler, James},
  title     = {Atkins' Physical Chemistry},
  edition   = {11},
  publisher = {Oxford University Press},
  year      = {2018},
  address   = {Oxford}
}

@article{hoeting1999bma,
  title={Bayesian model averaging: a tutorial},
  author={Hoeting, Jennifer A and Madigan, David and Raftery, Adrian E and Volinsky, Chris T},
  journal={Statistical Science},
  volume={14},
  number={4},
  pages={382--401},
  year={1999}
}

@article{spike_and_slab,
author = {Hemant Ishwaran and J. Sunil Rao},
title = {{Spike and slab variable selection: Frequentist and Bayesian strategies}},
volume = {33},
journal = {The Annals of Statistics},
number = {2},
publisher = {Institute of Mathematical Statistics},
pages = {730 -- 773},
year = {2005},
doi = {10.1214/009053604000001147}
}

@article{waterman_adami,
title = {Accelerated aging: Prediction of chemical stability of pharmaceuticals},
journal = {International Journal of Pharmaceutics},
volume = {293},
number = {1},
pages = {101-125},
year = {2005},
issn = {0378-5173},
doi = {https://doi.org/10.1016/j.ijpharm.2004.12.013},
author = {Kenneth C. Waterman and Roger C. Adami},
}

@article{king_84,
title = {Statistical Prediction of Drug Stability Based on Nonlinear Parameter Estimation},
journal = {Journal of Pharmaceutical Sciences},
volume = {73},
number = {5},
pages = {657-662},
year = {1984},
issn = {0022-3549},
doi = {https://doi.org/10.1002/jps.2600730517},
author = {Shang-Ying P. King and Min-Shya Kung and Ho-Leung Fung}

}

@article{bhangare2022,
title= {Systematic strategies for degradation kinetic 
study of pharmaceuticals: an issue of utmost 
importance concerning current stability analysis 
practices},
journal = {Journal of Analytical Science and Technology},
volume={13},
number={7},
pages={1-13},
year={2022},
doi={https://doi.org/10.1186/s40543-022-00317-6},
author={Bhangare, Dhiraj and Rajput, Niraj
            and Jadav, Tarang and Sahu, Amit Kumar
            and Tekade, Rakesh Kumar and Sengupta, Pinaki}
}

@article{clenet2014advanced,
  author  = {Cl{\'e}net, Didier and Imbert, Fr{\'e}d{\'e}ric and
             Probeck, Patricia and Rahman, Nausheen and Ausar, Salvador F.},
  title   = {Advanced kinetic analysis as a tool for formulation development
             and prediction of vaccine stability},
  journal = {Journal of Pharmaceutical Sciences},
  year    = {2014},
  volume  = {103},
  number  = {10},
  pages   = {3055--3064},
  doi     = {10.1002/jps.24117}
}

@article{skomski_25,
title = {Predictive stability in biopharmaceuticals and vaccines: Perspectives and recommendations towards accelerating patient access},
journal = {Journal of Pharmaceutical Sciences},
volume = {114},
number = {8},
pages = {103873},
year = {2025},
issn = {0022-3549},
doi = {https://doi.org/10.1016/j.xphs.2025.103873},
author = {Daniel Skomski and Andrea Ji and Drago Kuzman and Didier Clenet and Aaron Hieb and Scott W Roberts and Joe Berry and Christopher Lentes and Jos Weusten and Kirsten MacArthur and Amy St. Charles and Ben Ahlstrom and Sandra Auguste-Bowler and Leanne Chinn and Armin Boehrer and Shaoxin Feng and Chris Thompson and Bernard Francq and Christian Laue and Marie-Eve Bury and Adam Palmer Rauk and Thijs Cui and Matthew Scholfield and Michael Meleties and Yannick Kronimus and Kavitha Jakka and Matjaz Boncina and Pepijn Burgers and Elisabeth Krug and Edgardo Segarra and Jiewei Wu and Cavan Kalonia and Declan Lowney},
}

@article{dillon_et_al_24,
author = {Dillon, Michael and Xu, Jun and Thiagarajan, Geetha and Skomski, Daniel and Procopio, Adam},
title = {Predicting the Long-Term Stability of Biologics with Short-Term Data},
journal = {Molecular Pharmaceutics},
volume = {21},
number = {9},
pages = {4673-4687},
year = {2024},
doi = {10.1021/acs.molpharmaceut.4c00609},
    note ={PMID: 39121385}
}

@article{clenet,
title = {Accurate prediction of vaccine stability under real storage conditions and during temperature excursions},
journal = {European Journal of Pharmaceutics and Biopharmaceutics},
volume = {125},
pages = {76-84},
year = {2018},
issn = {0939-6411},
doi = {https://doi.org/10.1016/j.ejpb.2018.01.005},
url = {https://www.sciencedirect.com/science/article/pii/S0939641117308731},
author = {Didier Clénet}
}

\newpage

\setcounter{page}{1}
\setcounter{section}{0}
\setcounter{table}{0}
\setcounter{figure}{0}
\setcounter{equation}{0}
\renewcommand{\theHsection}{SIsection.\arabic{section}}
\renewcommand{\theHtable}{SItable.\arabic{table}}
\renewcommand{\theHfigure}{SIfigure.\arabic{figure}}
\renewcommand{\thepage}{S\arabic{page}}  
\renewcommand{\thesection}{S\arabic{section}}   
\renewcommand{\theequation}{S\arabic{equation}}   
\renewcommand{\thetable}{S\arabic{table}}   
\renewcommand{\thefigure}{S\arabic{figure}}
\renewcommand{\thetheorem}{S\arabic{theorem}}
\renewcommand{\thelemma}{S\arabic{lemma}}
\renewcommand{\theproposition}{S\arabic{proposition}}
\renewcommand{\thecorollary}{S\arabic{corollary}}
\renewcommand{\theequation}{S\arabic{equation}}

\appendix

\section*{Supplementary Material for "Bayesian modeling of predictive stability under generalized Arrhenius non-linear kinetics”}

\section{Full prior specification}

For all the models, we set $a_\sigma = b_\sigma = a_{Y_0} = b_{Y_0} = a_{k_{\text{ref}}} = b_{k_{\text{ref}}} =  5$ and $\mu_{k_{\text{ref}}} = -3$.\\
For the models assuming the Arrhenius equation (KKF-AE and SB-AE), we let $\mu_{E_a} = 9$ and $a_{E_a} = b_{E_a} = 5$, while for implementing the generalized Arrhenius equation framework, we empirically set $\mu_{E_a}$ to the posterior mean of the parameter $\log E_a$ under the model with Arrhenius dynamics. For the GAE models, we also set $a_K = b_K = 5$.\\
For the kinetic parameters, we let $\sigma_m^2 = 0.5$ and $\sigma_n^2 = 1$ for the SB model. For the SB-SAS model, we set $\sigma_{\mathrm{spike},m}^{2}=10^{-6}$, $\sigma_{\mathrm{slab},m}^{2}=0.5$, $\sigma_{\mathrm{spike},n}^{2}=10^{-4}$, and $\sigma_{\mathrm{slab},n}^{2}=1$.

\end{document}